\documentclass[fleqn,usenatbib]{mnras}

\usepackage{newtxtext,newtxmath}
\usepackage[T1]{fontenc}

\DeclareRobustCommand{\VAN}[3]{#2}
\let\VANthebibliography\thebibliography
\def\thebibliography{\DeclareRobustCommand{\VAN}[3]{##3}\VANthebibliography}

\usepackage{graphicx}
\usepackage{amsmath}
\usepackage{booktabs}
\usepackage{orcidlink}
\usepackage{xcolor}

\newcommand{\new}[1]{#1}
\newenvironment{revblock}{}{}

\title[V819 Herculis]{A joint astrometric, eclipse timing and radial velocity orbit for the hierarchical triple V819 Herculis}

\author[F. Akar \& \"O. Ba\c{s}t\"urk]{
Furkan Akar$^{\new{1}}$\thanks{E-mail: efurkanakar@gmail.com}\orcidlink{0000-0003-4419-2908}
and \"Ozg\"ur Ba\c{s}t\"urk$^{\new{2}}$\orcidlink{0000-0002-4746-0181}
\\
\new{$^{1}$\,Ankara University, Graduate School of Natural \& Applied Sciences, Astronomy \& Space Sciences Department,}\\
\new{Ziraat Mahallesi {\.I}rfan Ba\c{s}tu\u{g} Caddesi, D{\i}\c{s}kap{\i}, TR-06110 Alt{\i}nda\u{g} / Ankara, T\"urkiye}\\
\new{$^{2}$\,Ankara University, Faculty of Science, Astronomy \& Space Sciences Department, Degol cad., Emniyet mah.,}\\
\new{Yenimahalle, TR-06560, Ankara, T\"urkiye}\\
}

\date{Accepted 2026 November 01. Received 2026 October 07; in original form 2026 June 26}
\pubyear{\the\year{}}

\begin{document}
\label{firstpage}
\pagerange{\pageref{firstpage}--\pageref{lastpage}}
\maketitle

\begin{abstract}
We present a joint Bayesian orbital analysis of the hierarchical triple system V819 Herculis using ground-based astrometry, radial velocities (RV), and eclipse timing variations (ETV). By coupling these complementary data sets within a physically consistent Keplerian framework, we obtain a self-consistent three-dimensional solution for the outer orbit and the dynamical properties of the system. The outer orbit is found to have a period of \new{$P_3 = 2016.97 \pm 0.27$\,d} and a high eccentricity of \new{$e_3 = 0.6800 \pm 0.0016$}, with the same orbital geometry simultaneously reproducing the sky-plane astrometry, the LiTE signal in the eclipse timings, and the tertiary radial velocities. \new{The joint solution determines the component masses dynamically from these data alone, yielding $M_{12} = 2.721 \pm 0.055\,M_\odot$ for the inner binary and $M_3 = 1.954 \pm 0.070\,M_\odot$ for the tertiary,} together with a LiTE amplitude of \new{$A_{\rm LiTE} = 0.0104$\,d} and an RV semi-amplitude of \new{$K_3 = 18.46$\,km\,s$^{-1}$}. \new{The joint fit also yields an orbital parallax of $\pi_{\rm orb} = 14.08 \pm 0.13$\,mas, consistent with Gaia DR3, Gaia DR2 and the HIPPARCOS New Reduction.} The derived tertiary mass is higher than expected for a standard G8-type main-sequence star, suggesting that the tertiary component may be in an advanced evolutionary stage. The resulting outer orbit is well constrained\new{, and the system is strongly hierarchical, with an outer period about 900 times the inner one}. Finally, we provide predicted astrometric positions for the tertiary component for five years beyond 2026.5 to guide future high-precision observations.
\end{abstract}


\section{Introduction}
\label{sec:intro}

Understanding the architecture of multiple stellar systems requires observations that probe orbital motion in more than one dimension. Astrometry traces the projected motion of the photocentre on the plane of the sky, eclipse timing variations (ETV) of an inner eclipsing binary encode the light-travel-time delay as the binary revolves around the system's centre of mass, and radial velocity (RV) measurements probe the line-of-sight component of barycentric motion. Since each of these techniques provides information in a different
dimension, their combined analysis allows us to recover the nature of the motion far more completely than any approach based on a single dimension. Each technique in isolation, however, is subject to its own geometric degeneracies: astrometry cannot resolve the distance scale without an independent parallax, ETV signals constrain only the projected semi-major axis along the line of sight, and RV measurements are
insensitive to the inclination of the outer orbit. When all three data streams are modelled simultaneously within a single self-consistent framework, these degeneracies lift in a complementary fashion. The sky-plane orbit fixes the angular scale and the inclination, the light-time effect constrains the line-of-sight semi-major axis, and the RV curve fixes the physical velocity amplitude, all tied to the same Keplerian geometry. The result is a three-dimensional outer-orbit solution in which the parallax, component masses, and outer-orbit inclination are constrained within the joint model rather than fixed to external values. Outer companions of eclipsing binaries are particularly compelling targets for this approach: their orbital characteristics can only be fully recovered by exploiting all available projections of the motion simultaneously, and it is precisely the determination of the orbital and physical properties of such additional bodies that is crucial for understanding the nature of multi-stellar systems.

V819 Herculis (HR 6469 = HD 157482 = HIP 84949) is a system exceptionally suitable for such a combined analysis. It consists of an evolved giant (G7~III--IV, designated A) in an eccentric 5.5-year outer orbit around a close pair of F-type main-sequence stars (Ba and~Bb) that eclipse each other on a 2.23-day circular orbit \citep{scarfe1994, vanHamme1994}. All three observational ingredients are available with substantial historical depth: four decades of interferometric astrometry covering multiple orbital cycles, a 17-year spectroscopic monitoring campaign from four observatories \citep{scarfe1994}, and eclipse timings spanning more than 40 years including five sectors of \textit{Transiting Exoplanet Survey Satellite} (TESS) data. Crucially, the bright and well-resolved nature of the system means that these data sets have been accumulated independently to a level of precision rarely available for triple stars, making V819 Her one of the few systems where a fully coupled three-dimensional Bayesian solution is feasible.

The observational history of the system is long and rich. The eclipsing nature of the inner pair was discovered by \citet{boyd1985}, who also identified the quasi-periodic photometric variability of the outer giant attributed to starspot activity. \citet{vanHamme1994} solved the light curve with the Wilson-Devinney code \citep{wilsondev1971, wilsondev1973}, determining the third-light contribution and spectral classifications of F2 V and F8 V for the two eclipsing components. \citet{wasson1994} collected ninety eclipse timings over a decade and demonstrated a coherent light-travel-time effect consistent with the spectroscopic outer orbit. \citet{scarfe1994} combined radial velocities from the David Dunlap, McDonald, Kitt Peak, and Dominion Astrophysical Observatories with speckle interferometry to derive orbital elements and physical properties for all three components, obtaining an orbital parallax of $\pi_{\rm orb} = 14.5 \pm 0.2$ mas. High-precision differential astrometry from the PHASES programme at the Palomar Testbed Interferometer then transformed the precision of the outer orbit. \citet{muterspaugh2006} resolved the photocentre wobble of the inner Ba--Bb pair at the $10$--$20\,\mu$as level and determined the mutual inclination between the inner and outer orbital planes to be $23.6 \pm 4.9\degr$, representing the sixth unambiguous determination of this quantity in any hierarchical triple at the time. An updated solution by \citet{muterspaugh2008} refined the component masses to $M_{12} = 2.560 \pm 0.067\,M_\odot$ and $M_3 = 1.799 \pm 0.098\,M_\odot$. \citet{obrien2011} used the CHARA Array to resolve the inner Ba--Bb orbit through separated fringe packets, confirming the mutual inclination at $33.5 \pm 9.3\degr$ and estimating the system age at $1.9 \pm 1.1$~Gyr. \citet{zasche2014} combined the full interferometric data set with all available eclipse timings in a simultaneous astrometry-plus-LiTE framework, deriving a geometric distance of $68.8 \pm 1.8$ pc.

\begin{revblock}
More recently, high-angular-resolution optical speckle observations of V819\,Her were obtained with the `Alopeke instrument on the Gemini North telescope by \citet{howell2024}. Using Multi-Frame Blind Deconvolution, they resolved the close eclipsing pair from the brighter tertiary component and reported the tentative detection of a faint additional source.
\end{revblock}

Despite this wealth of data, no prior study has coupled all three observational channels for V819\,Her within a single probabilistic model. Such a combination is uncommon among hierarchical triples, because it requires resolved astrometry, long-term radial velocities, and eclipse timings that all cover the same outer orbit with sufficient precision. The analyses of \citet{scarfe1994} and \citet{muterspaugh2008} combined astrometry and RV but treated the eclipse timings separately; \citet{zasche2014} combined astrometry and the light-travel-time signal but omitted the radial velocities of the outer orbit. As a result, residual correlations between the parallax, the outer orbital inclination, and the component masses have never been fully resolved for this system, and the component masses in particular remain sensitive to which data combinations and priors are adopted. The arrival of \textit{TESS} photometry in Sectors 25, 26, 52, 53, and~80 has substantially extended and densified the eclipse timing baseline, strengthening the constraining power of the light-time effect beyond what was available to \citet{zasche2014}.

In this paper we present a joint Bayesian analysis of astrometric, ETV, and RV data for V819~Her within a unified Keplerian framework. \new{The angular semi-major axis, the parallax, the outer period and the mass ratio are sampled simultaneously, so that the component masses are determined dynamically from the data.} This coupling allows the orbital inclination and distance scale to be constrained dynamically, rather than being fixed to external catalogue values. In particular, the parallax is treated as a free parameter with an informative Gaia DR3 prior, so that the resulting \new{orbital} parallax can be assessed against the HIPPARCOS New Reduction \citep{vanleeuwen2007} and the Gaia DR2 and DR3 catalogues \citep{gaiadr2, gaiadr3}. This comparison is especially relevant for V819\,Her, since its Gaia DR3 solution has a high Renormalised Unit Weight Error (RUWE), indicating that the standard single-star astrometric solution is likely affected by unresolved orbital motion \citep{lindegren2018, lindegren2021}.

The remainder of the paper is organised as follows. Section~\ref{sec:data} describes the three observational data sets. Section \ref{sec:theory} presents the theoretical framework for each observable. The joint model, priors, and sampling strategy are detailed in Section \ref{sec:jointmodel}, and the results and their astrophysical implications are discussed in Section \ref{sec:discussion}.

\section{Data}
\label{sec:data}
\begin{figure*}
    \centering
    \includegraphics[width=\textwidth]{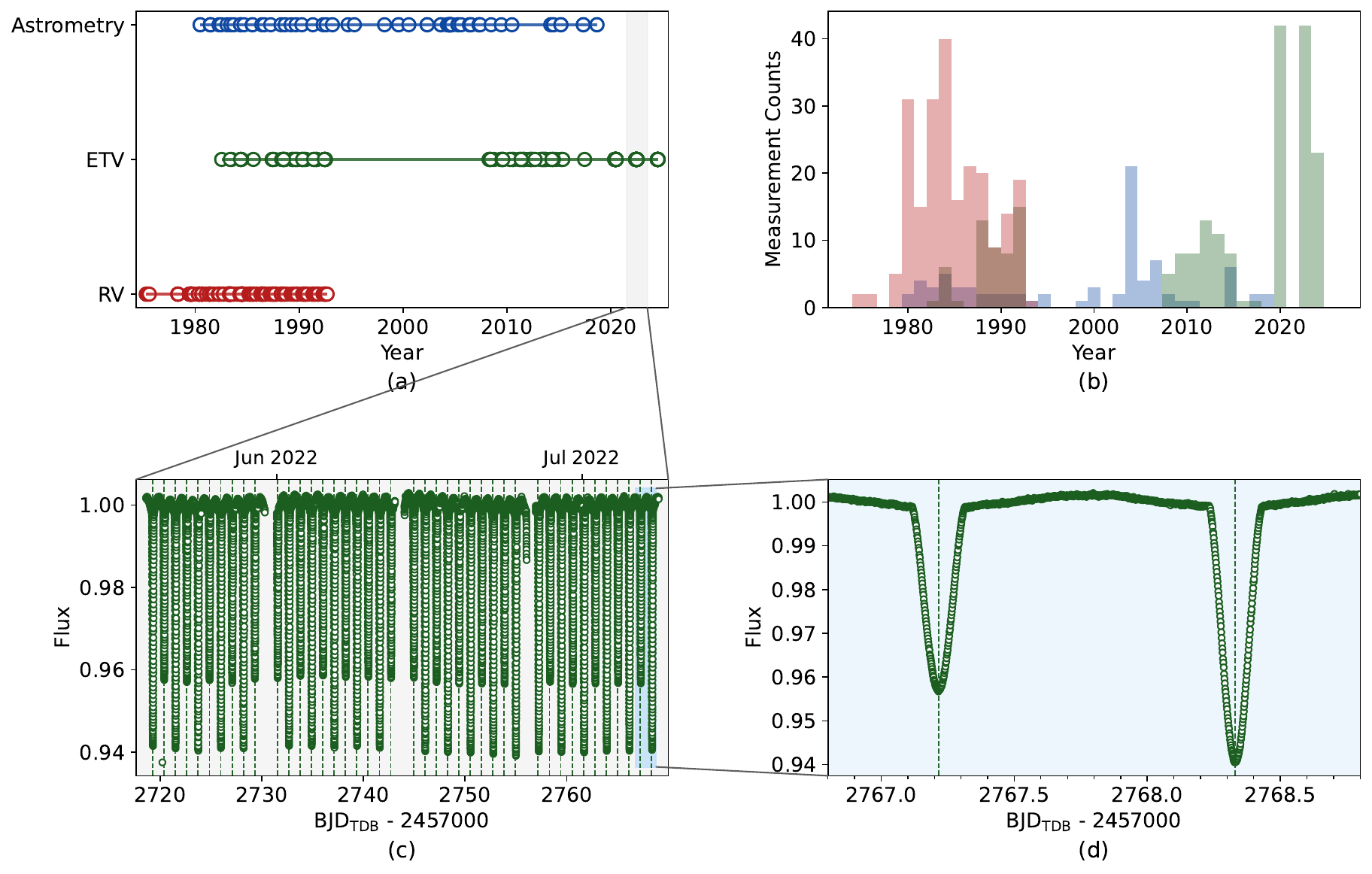}
    \caption{Summary of observational data used for V819 Her.
    Panel (a) shows the temporal distribution of astrometric (blue), ETV (green), and RV (red) data;
    panel (b) presents the corresponding measurement counts;
    panels (c) and (d) display segments of the TESS light curve from Sectors 52--53, including a close-up of a single orbital cycle.}
    \label{fig:data_summary}
\end{figure*}

This study combines three independent observational data sets of the V819 Her system: astrometric measurements, eclipse timing variations (ETV), and archival radial velocities (RV). Each data \new{set spans more than a decade} and originates from different observational techniques. A summary of the temporal coverage and measurement density for all data types is presented in Fig.~\ref{fig:data_summary}. \new{All epochs were converted to the barycentric dynamical time scale (BJD$_{\rm TDB}$): the Besselian epochs of the astrometric measurements and the heliocentric Julian dates of the published eclipse timings and radial velocities.}

\subsection{Astrometry}
\label{subsec:astrometry}

Astrometric measurements of the V819 Her system were obtained from the Washington Double Star Catalogue (WDS\footnote{\url{https://www.astro.gsu.edu/wds}}), which is maintained and regularly updated by the United States Naval Observatory (USNO). The catalogue provides quantitative and qualitative parameters for each observation, such as the observation epoch, position angle (\(\theta\)) in degrees, angular separation (\(\rho\)) in arcseconds, observing technique, telescope aperture, and various notes about the system. \new{The WDS lists 129 measurements with complete information for epoch, \(\rho\), and \(\theta\); their epochs are given as fractional Besselian years.}

\begin{revblock}
For the long-baseline interferometric measurements of the PHASES programme, we use the values of the PHASES archive \citep{muterspaugh2010}. Following \citet{muterspaugh2008, muterspaugh2010}, we exclude ten PTI measurements taken during inner-binary eclipses, when the photocentre of the inner binary is displaced, and two measurements flagged as outliers in the PHASES archive. The two speckle measurements of \citet{horch2021} at epoch 2018.65 have position angles that differ by $180\degr$ from the orbit traced by all other data, the well-known quadrant ambiguity of speckle interferometry, to which pairs with small magnitude differences ($\Delta m = 0.78$ and $0.95$\,mag here) are particularly prone \citep{horch2021}; their position angles were corrected accordingly. The final astrometric data set contains 83 measurements spanning 38 years (1980.5--2018.7).
\end{revblock}

The WDS lists three observing techniques for V819~Her: \textit{speckle interferometry}, \textit{CHARA speckle}, and the \textit{Palomar Testbed Interferometer (PTI)}. These labels correspond to the observing methods reported in the catalogue and were grouped into two broader categories according to their measurement principles: \textit{speckle interferometric} and \textit{long-baseline interferometric} observations. All long-baseline interferometric data include reported uncertainties, whereas \new{51} of the speckle measurements do not. For these entries, missing uncertainties were assigned by adopting \new{the largest uncertainties published for the speckle measurements}, corresponding to \(\sigma_\theta = 2^{\circ}\) for position angle and \(\sigma_\rho = 0^{\prime\prime}.003\) for angular separation\new{; the same values were applied to the CHARA speckle measurements, for which no uncertainties are reported}. To account for potential underestimation of the formal errors and to capture additional unmodelled scatter, additive jitter terms were also included in both position angle and separation during the orbital modelling.

\subsection{Eclipse Timing}
\label{subsec:photometry}

Eclipse timing data for V819 Her cover the time span from 1982 to 2024. Previously published times of minima were compiled from the literature, while new measurements were derived from photometric observations obtained by the TESS. The system was observed by TESS in Sectors 25, 26, 52, 53, and 80, primarily with Camera 2 and with Camera 3 in the final sector, using a cadence of 120 s. In total, \new{215} eclipse timings were collected for this analysis. In the compiled literature data, when multiple minima corresponding to the same orbital cycle were reported in different photometric filters, they were combined into a single representative value by calculating their weighted mean.

The TESS photometric data were retrieved and preprocessed using \texttt{lcm}\footnote{\url{https://github.com/efurkanakar/lcm}}, a light-curve management tool that integrates multiple data-handling utilities, including the \texttt{astroquery} package \citep{astroquery2019}. We used the Simple Aperture Photometry (SAP) light curves and normalized each light curve by its median flux. Times of minima were then determined from the processed light curves using the \texttt{Xtrema} software \citep{xtrema2015}, which applies the classical Kwee--van Woerden method \citep{kwee1956}. The resulting minima were combined with the published values to produce the final set of eclipse timings used in this study.

\subsection{Radial Velocity}
\label{subsec:rv}

Radial velocity measurements of the long-period tertiary component in the V819 Her system were adopted from \citet{scarfe1994}, who conducted a spectroscopic monitoring campaign between 1975 and 1992. The observations were obtained at the David Dunlap Observatory (DDO), McDonald Observatory (McD), Kitt Peak National Observatory (KPNO), and Dominion Astrophysical Observatory (DAO), yielding a total of 226 individual measurements. \new{This campaign remains the source of the epoch radial velocities used in the later orbital analyses of the system \citep{muterspaugh2006, muterspaugh2008}, while Gaia DR3 provides only summary statistics of its RVS velocities, which we compare with our orbit in Section~\ref{sec:discussion}.}

Uncertainties for the individual observatories were assigned in accordance with the measurement precisions discussed by \citet{scarfe1994}. Specifically, values of 2~km~s$^{-1}$, 0.5~km~s$^{-1}$, and 1~km~s$^{-1}$ were adopted for the DDO, McDO-KPNO, and DAO data, respectively.

\section{Orbital Geometry and Theoretical Framework}
\label{sec:theory}

\subsection{Astrometry}
\label{subsec:astrometry_theory}

The orbital motion of the tertiary component around the eclipsing binary can be described using the classical two-body Keplerian formulation. The apparent motion observed on the sky is a geometric projection of this orbit onto the plane of the celestial sphere. The tertiary’s orbit can be characterized by the standard set of Keplerian elements:
\begin{equation}
\Theta_{\mathrm{orb}} =
(a_{\mathrm{ang}},\, e_3,\, i_3,\, \Omega_3,\, \omega_3,\, P_3,\, T_{\mathrm{per}}),
\label{eq:kepler_elements}
\end{equation}
where $a_{\mathrm{ang}}$ is the angular semi–major axis, $e_3$ the eccentricity, $i_3$ the inclination, $\Omega_3$ the longitude of the ascending node, $\omega_3$ the argument of periastron, $P_3$ the orbital period, and $T_{\mathrm{per}}$ the epoch of periastron passage.

At any given time, the position of the tertiary on its elliptical orbit is described by the mean, eccentric, and true anomalies. The mean anomaly is defined as
\begin{equation}
M(t) = n\,(t - T_{\mathrm{per}}), \qquad n = \frac{2\pi}{P_3},
\end{equation}
The relation between the mean anomaly $M$ and the eccentric anomaly $E$ is expressed through Kepler’s equation,
\begin{equation}
M = E - e_3\sin E,
\end{equation}
which is solved numerically for $E$. The true anomaly $\nu$, representing the angular distance of the body from periastron as seen from the focus, is then computed as
\begin{equation}
\nu = 2 \arctan \!\left[\sqrt{\frac{1+e_3}{1-e_3}}\tan\frac{E}{2}\right],
\end{equation}
and the instantaneous orbital radius follows from
\begin{equation}
r = a_{\rm rel}\,(1 - e_3\cos E).
\end{equation}

In the orbital plane, the barycentric position of the tertiary relative to the binary can be written as
\begin{equation}
(x_{\mathrm{orb}},\, y_{\mathrm{orb}},\, 0) =
(r\cos\nu,\, r\sin\nu,\, 0).
\end{equation}

To obtain the apparent coordinates on the plane of the sky, these positions are projected using the standard rotation by the orbital angles: inclination
($i_3$), argument of periastron ($\omega_3$), and longitude of the ascending
node ($\Omega_3$):
\begin{align}
X &= r\,[\cos\Omega_3\cos(\omega_3+\nu)
       - \sin\Omega_3\cos i_3\,\sin(\omega_3+\nu)], \\
Y &= r\,[\sin\Omega_3\cos(\omega_3+\nu)
       + \cos\Omega_3\cos i_3\,\sin(\omega_3+\nu)], \\
Z &= -r\,\sin i_3\,\sin(\omega_3+\nu),
\end{align}
where $X$ is oriented northward, $Y$ eastward, and $Z$ is directed along the
line of sight toward the observer.

These projection equations follow the astrometric conventions implemented in
the \texttt{exoplanet} package (\citealt{Foreman-Mackey2021}).

The directly observable quantities in astrometric data are the angular separation ($\rho$) and position angle ($\theta$), which are related to these projected coordinates through
\begin{align}
\rho &= \sqrt{X^2 + Y^2}, \\
\theta &= \arctan2(Y, X).
\end{align}

These relations form the complete geometric model linking the orbital elements of the tertiary’s orbit to the sky-plane observables. If a trigonometric parallax is available, the angular semi-major axis measured on the sky can be converted to a physical scale, and Kepler’s third law then yields the total mass. Using the distance $d = 1/\pi$ (with $\pi$ in arcsec and $d$ in pc), the relative semi-major axis in AU is
\begin{equation}
a_{\rm rel}\,[{\rm AU}] \;=\; a_{\rm rel}\,[{\rm arcsec}] \times d\,[{\rm pc}],
\end{equation}
where $a_{\rm rel}=a_3+a_{12}$ is the two-body relative semi-major axis of the tertiary and the binary. With $P_3$ in years and $a_{\rm rel}$ in AU, the system’s total mass follows as
\begin{equation}
M_{\rm tot} \; [M_\odot]=\; M_{12}+M_3 \;=\; \frac{a_{\rm rel}^3}{P_3^2},
\label{eq:kepler3}
\end{equation}
where $M_{12}$ and $M_3$ are the total mass of the inner eclipsing binary and the mass of the tertiary component, respectively.

\begin{figure}
    \centering
    \includegraphics[width=\columnwidth]{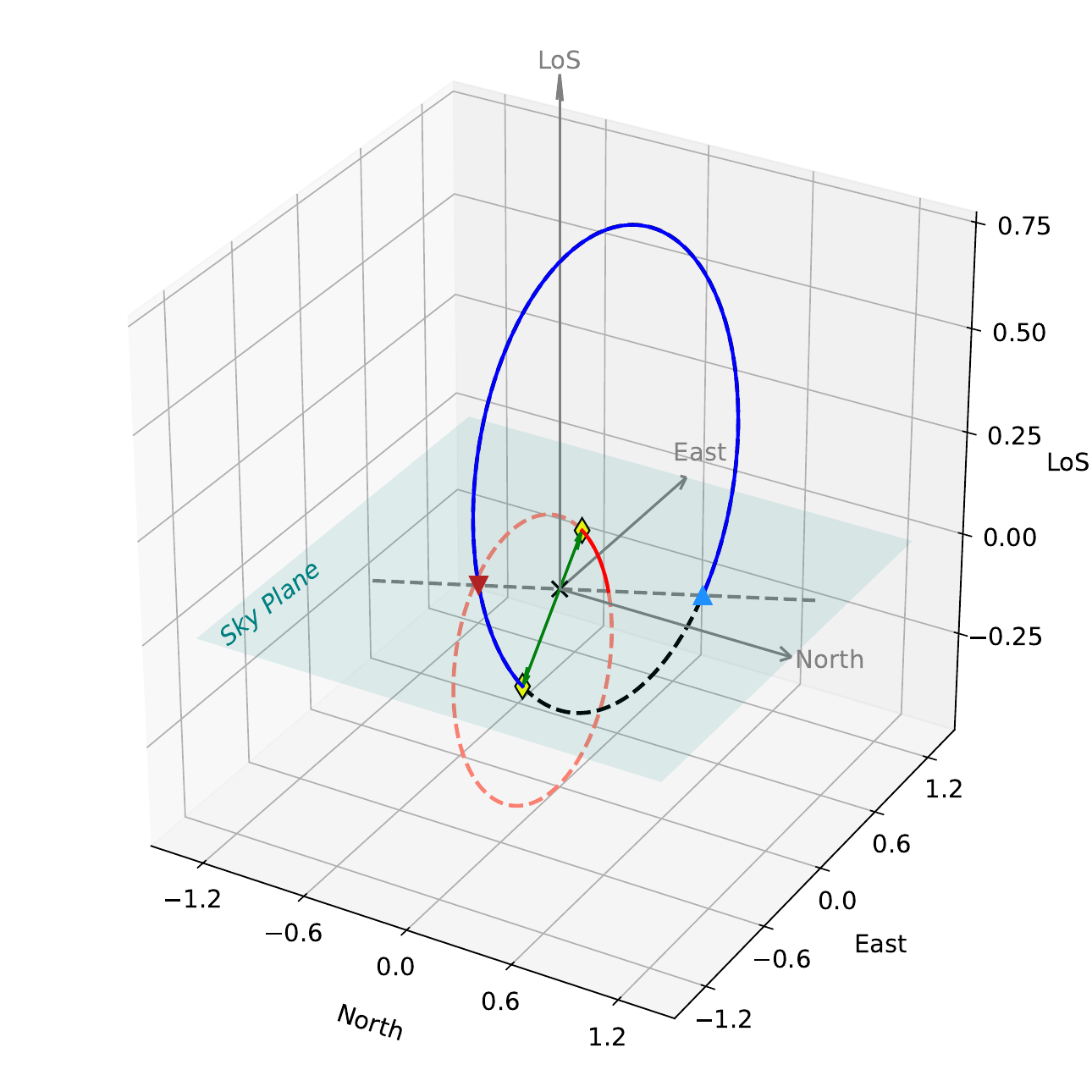}
    \caption{
    Schematic 3D geometry of the system used to define periastron directions and node conventions.
    The cyan plane indicates the sky plane, oriented within the North–East directions and the Line of Sight (LoS) axis shown in gray. The dashed red and solid blue ellipses represent the barycentric orbits of the binary and the tertiary components around their common centre of mass (black cross), respectively. Periastron points are marked with yellow diamonds, and short red and blue arcs illustrate the arguments of periastron: \new{$\omega_{3}=240^\circ$ for the tertiary and $\omega_{12}=60^\circ$ for the binary’s barycentric motion}. The green dashed line connects the two periastron directions, indicating that the two arguments of periastron differ by $180^\circ$. The ascending and descending nodes are shown as blue and red triangles, while the gray dashed line marks the line of nodes.
    }
    \label{fig:omega_relation}
\end{figure}

\subsection{Light–Time Effect from Eclipse Timings}
\label{subsec:lite_theory}

Variations in the observed times of eclipses in a binary system can arise when the binary is in orbital motion around a common centre of mass with a third body. As the binary moves toward or away from the observer along its outer orbit, the light-travel time between the system and the Earth changes periodically, producing apparent delays or advances in the observed times of minimum light, a phenomenon known as the Light–Time Effect (LiTE).

The observed times of minima can be expressed as a linear ephemeris combined with the LiTE correction:
\begin{equation}
T_{\mathrm{min}}(\new{E_{\rm cyc}}) = T_0 + \new{E_{\rm cyc}}\,P_{12} + \Delta t_{\mathrm{LiTE}},
\end{equation}
where $T_0$ is the reference epoch, \new{$E_{\rm cyc}$ the cycle number,} $P_{12}$ the orbital period of the binary, and $\Delta t_{\mathrm{LiTE}}$ the light-travel-time delay induced by the barycentric motion of the binary system.

Following the classical formulation by \citet{irwin1952}, the LiTE delay is given by
\begin{equation}
\Delta t_{\mathrm{LiTE}} =
\frac{a_{12}\sin i_3}{c}
\left[
\frac{(1-e_3^2)\sin(\nu+\omega_{12})}{1+e_3\cos\nu}
+ e_3\sin\omega_{12}
\right],
\end{equation}
where $a_{12}$ is the semi-major axis of the binary’s barycentric orbit, $i_3$ and $e_3$ are the inclination and eccentricity of the tertiary’s orbit, $\omega_{12}$ is the argument of periastron of the binary’s barycentric motion, and $c$ is the speed of light. The true anomaly $\nu$ corresponds to the orbital phase of the tertiary’s orbit at the time of observation. The semi-amplitude of this light-time signal is defined as:
\begin{equation}
A_{\rm LiTE} = \frac{a_{12}\sin i_3}{c}.
\end{equation}

In the adopted geometric convention, the argument of periastron of the binary’s barycentric orbit ($\omega_{12}$) and that of the tertiary component ($\omega_3$) are related by $\omega_{12} = \omega_3 + 180^\circ$, since both refer to the same line of apsides but measured from opposite orbital foci. However, the literature often uses inconsistent definitions of these angles or omits them entirely, leading to ambiguity in interpreting periastron directions. To avoid such confusion, the adopted orientation and angle definitions are illustrated in Fig.~\ref{fig:omega_relation}.

\subsection{Radial Velocity of the Tertiary Component}
\label{subsec:rv_theory}

The radial velocity of the tertiary component reflects its orbital motion around the common centre of mass with the binary. At any time, the instantaneous line-of-sight velocity of the tertiary is given by the classical Keplerian expression:
\begin{equation}
V_{r_3} = \gamma + K_{3}\,[\cos(\nu+\omega_{3}) + e_3\cos\omega_{3}],
\end{equation}
where $\gamma$ is the systemic velocity of the triple system, $\nu$ is the true anomaly angle for the tertiary, $\omega_{3}$ is the argument of periastron, and $e_3$ the eccentricity.

The velocity semi-amplitude of the tertiary is
\begin{equation}
K_{3} =
\frac{2\pi\,a_{3}\sin i_3}{P_3\,\sqrt{1-e_3^2}},
\end{equation}
where $a_3$, $i_3$, and $P_3$ are the semi-major axis, inclination, and orbital period of the tertiary’s orbit, respectively.

In this study, the available radial velocity measurements correspond to the tertiary component, thus tracing its motion directly rather than the reflex motion of the binary.

Taken together, the astrometric, eclipse timing, and radial velocity observables represent three independent projections of the same Keplerian motion: the astrometry constrains the orbit on the sky plane, the LiTE signal measures the light-travel time along the line of sight, and the radial velocities trace the instantaneous barycentric motion of the tertiary. In the following section, these \new{data sets} are combined within a single probabilistic framework to obtain a consistent and physically coherent orbital solution.

\begin{table}
\centering
\caption{Parameters of the joint model and the observational data sets that constrain them. A cross indicates that the parameter is constrained either directly by the corresponding observable or indirectly through the physical coupling imposed by the joint model. The symbol (d) denotes quantities deterministically derived from sampled parameters\new{; it is placed under the observable whose model uses the quantity, and under all three for the masses, which enter no observable directly}.}
\label{tab:constraints_table}
\setlength{\tabcolsep}{6pt}
\renewcommand{\arraystretch}{1.2}
\begin{tabular}{l c c c c}
\toprule
Parameter & Symbol & Astrometry & ETV & RV \\
\midrule
\multicolumn{5}{c}{\textit{Free Parameters}} \\
\midrule
Angular semi–major axis & $a_{\rm ang}$        & $\times$ & $\times$ & $\times$ \\
Orbital period          & $P_3$                & $\times$ & $\times$ & $\times$ \\
Eccentricity            & $e_3$                & $\times$ & $\times$ & $\times$ \\
Inclination             & $i_3$                & $\times$ & $\times$ & $\times$ \\
Argument of periastron  & $\omega_3$           & $\times$ & $\times^\dagger$ & $\times$ \\
Longitude of asc. node  & $\Omega_3$           & $\times$ &  &  \\
Periastron epoch        & $T_{\rm per}$        & $\times$ & $\times$ & $\times$ \\
Parallax                & $\pi$                & $\times$ & $\times$ & $\times$ \\
\new{Mass ratio}        & \new{$q_3$} &  & \new{$\times$} & \new{$\times$} \\
Epoch Correction        & $dT$                 &  & $\times$ &  \\
Binary Period Corr.     & $dP$                 &  & $\times$ &  \\
Systemic velocity       & $\gamma$             &  &  & $\times$ \\
\new{Separation jitter}   & \new{$\sigma_{\rho,{\rm jit}}$}   & \new{$\times$} &  &  \\
\new{Position-angle jitter} & \new{$\sigma_{\theta,{\rm jit}}$} & \new{$\times$} &  &  \\
\new{Timing jitter}       & \new{$\sigma_{{\rm OC},{\rm jit}}$} &  & \new{$\times$} &  \\
\new{RV jitter}           & \new{$\sigma_{{\rm RV},{\rm jit}}$} &  &  & \new{$\times$} \\
\midrule
\multicolumn{5}{c}{\textit{Deterministically Derived Parameters}} \\
\midrule
\new{Total mass}        & \new{$M_{\rm tot}$}  & \new{(d)} & \new{(d)} & \new{(d)} \\
\new{Mass of inner binary} & \new{$M_{12}$}    & \new{(d)} & \new{(d)} & \new{(d)} \\
\new{Mass of tertiary}  & \new{$M_3$}          & \new{(d)} & \new{(d)} & \new{(d)} \\
LiTE amplitude          & $A_{\rm LiTE}$       &  & (d) &  \\
RV amplitude            & $K_3$                &  &  & (d) \\
\bottomrule
\end{tabular}
\vspace{4pt}
\parbox{0.94\linewidth}{\small $^\dagger$ The ETV model uses the geometrical relation \new{$\omega_{12}=\omega_3+180^\circ$} between the binary and tertiary periastron directions.}
\end{table}

\section{Combined Analysis}
\label{sec:jointmodel}

\subsection{Model Architecture and Physical Coupling}
\label{subsec:architecture}

To obtain a consistent orbital solution for V819 Herculis, we simultaneously analysed the astrometric, ETV, and RV \new{data sets} within a single Bayesian framework. Instead of modelling the sky-plane motion, timing variations, and radial velocities independently, we coupled all three observables through a shared Keplerian orbit.

\begin{table*}
\centering
\caption{Summary of prior distributions assigned to the free parameters in the joint orbital model. $\mathcal{U}$, $\mathcal{N}$, and $\mathcal{TN}$ denote Uniform, Normal, and Truncated Normal distributions, respectively.}
\label{tab:priors_joint_final_units}
\setlength{\tabcolsep}{10pt}
\renewcommand{\arraystretch}{1.1}
\small
\begin{tabular}{lllll}
\toprule
Parameter & Symbol & Units & Prior Type & Hyperparameters / Bounds \\
\midrule
\multicolumn{5}{l}{\textit{\new{Outer Orbital Parameters}}} \\
\midrule
Angular semi–major axis     & $a_{\rm ang}$  & arcsec        & \new{$\mathcal{TN}$}   & $\mu=0.075$, $\sigma=0.10$\new{, lower bound $0$} \\
Orbital period (log scale)  & $\log P_3$     & $\log({\rm d})$& $\mathcal{N}$   & $\mu=\log(2000)$, $\sigma=\log(1.25)$ \\
Eccentricity                & $e_3$          & —             & \new{Beta}  & \new{$\alpha=0.867$, $\beta=3.03$} \\
Inclination (via cosine)    & $\cos i_3$     & —             & $\mathcal{U}$   & bounds $[-1,\,1]$ \\
Angle reparameterisation    & $p,\,m$        & rad           & \new{$\mathcal{U}$}   & \new{bounds $[-\pi,\,\pi)$} \\
Periastron phase            & $\mathrm{phase}$& rad          & \new{$\mathcal{U}$}   & \new{bounds $[-\pi,\,\pi)$}$^\dagger$ \\
\midrule
\multicolumn{5}{l}{\textit{Physical and System Properties}} \\
\midrule
Observed parallax           & $m_{\pi}$      & mas           & $\mathcal{TN}$  & $\mu=13.815$, $\sigma=0.496$, bounds $[0,\,50]$ \\
\new{Mass ratio}            & \new{$q_3$}    & \new{—}       & \new{$\mathcal{U}$} & \new{bounds $[0.1,\,2]$} \\
Systemic velocity           & $\gamma$       & ${\rm km\,s^{-1}}$ & $\mathcal{N}$ & $\mu=-3.5$, $\sigma=5.0$ \\
\midrule
\multicolumn{5}{l}{\textit{Ephemeris Corrections}} \\
\midrule
Binary Epoch correction     & $dT$           & d             & $\mathcal{N}$   & $\mu=0$, $\sigma=5\times10^{-3}$ \\
Binary Period correction    & $dP$           & d             & $\mathcal{N}$   & $\mu=0$, $\sigma=1\times10^{-5}$ \\
\midrule
\multicolumn{5}{l}{\textit{\new{Jitter Terms}}} \\
\midrule
Astrometric (separation)    & \new{$\log \sigma_{\rho,{\rm jit}}$}  & $\log({\rm arcsec})$ & $\mathcal{N}$ & $\mu=\log[\mathrm{median}(\sigma_\rho)]$, $\sigma=1.0$ \\
Astrometric (pos. angle)    & \new{$\log \sigma_{\theta,{\rm jit}}$}& $\log({\rm rad})$    & $\mathcal{N}$ & $\mu=\log[\mathrm{median}(\sigma_\theta)]$, $\sigma=1.0$ \\
Radial velocity             & $\log \sigma_{{\rm RV}, {\rm jit}}$ & $\log({\rm km\,s^{-1}})$ & $\mathcal{N}$ & $\mu=\log[\mathrm{median}(\sigma_{\rm RV})]$, $\sigma=1.0$ \\
Eclipse timing              & $\log \sigma_{{\rm OC}, {\rm jit}}$ & $\log({\rm d})$          & $\mathcal{N}$ & $\mu=\log[\mathrm{median}(\sigma_{\rm OC})]$, $\sigma=1.0$ \\
\bottomrule
\end{tabular}
\vspace{3pt}
\begin{flushleft}
\vbox{\footnotesize $^\dagger$ \new{$T_{\rm per}$ is derived as $T_{\rm per}=T_{\rm ref}+P_3\,\mathrm{phase}/(2\pi)$, with the reference epoch $T_{\rm ref} = 2452621.59$ (2002.95).}}
\end{flushleft}
\end{table*}

The Keplerian kinematics and geometric projections described in Section \ref{sec:theory} were numerically evaluated using the \texttt{exoplanet} package \citep{Foreman-Mackey2021}. While the foundational kinematics are described by the classical elements in equation~(\ref{eq:kepler_elements}), our Bayesian model samples a reparameterised and expanded vector, $\Theta$, to optimize MCMC efficiency and enforce physical coupling. The complete set of free parameters, including the \new{mass ratio}, parallax, systemic velocity, ephemeris corrections, and \new{data set}-specific jitter terms, along with the specific observables constraining them, is summarised in Table \ref{tab:constraints_table}.

\begin{revblock}
The strict dynamical consistency of the joint model is achieved by tying the component masses to the orbit through Kepler's third law, rather than treating the velocity semi-amplitude ($K_3$), the LiTE amplitude ($A_{\rm LiTE}$) or the masses themselves as independent free variables. At each MCMC step, the angular semi-major axis ($a_{\rm ang}$), the parallax ($\pi$), the outer period ($P_3$) and the mass ratio ($q_3 = M_3/M_{12}$) are sampled. The physical relative semi-major axis is then uniquely determined as $a_{\rm rel} = a_{\rm ang}/\pi$, with both $a_{\rm ang}$ and $\pi$ expressed in arcseconds, and the total mass follows from equation~(\ref{eq:kepler3}),
\begin{equation}
M_{\rm tot} = \frac{4\pi^2 a_{\rm rel}^3}{G P_3^2}, \qquad M_{12} = \frac{M_{\rm tot}}{1+q_3}, \qquad M_3 = q_3\,M_{12}.
\end{equation}
The relative orbit is partitioned between the inner binary and the tertiary component around their common centre of mass:
\end{revblock}

\begin{equation}
a_{12} = a_{\rm rel} \frac{M_3}{M_{12} + M_3}, \qquad a_3 = a_{\rm rel} \frac{M_{12}}{M_{12} + M_3}.
\end{equation}
The required observational semi-amplitudes ($K_3$ and $A_{\rm LiTE}$) are subsequently derived as deterministic functions of these partitioned physical scales, using the standard relations defined in Sections \ref{subsec:lite_theory} and \ref{subsec:rv_theory}. Through this formulation, the ETV and RV \new{data sets} actively contribute to constraining the overall geometry of the system, including the orbital inclination ($i_3$) and the absolute spatial scale. \new{No external mass information enters the fit.}

\begin{figure*}
    \centering
    \includegraphics[width=\textwidth]{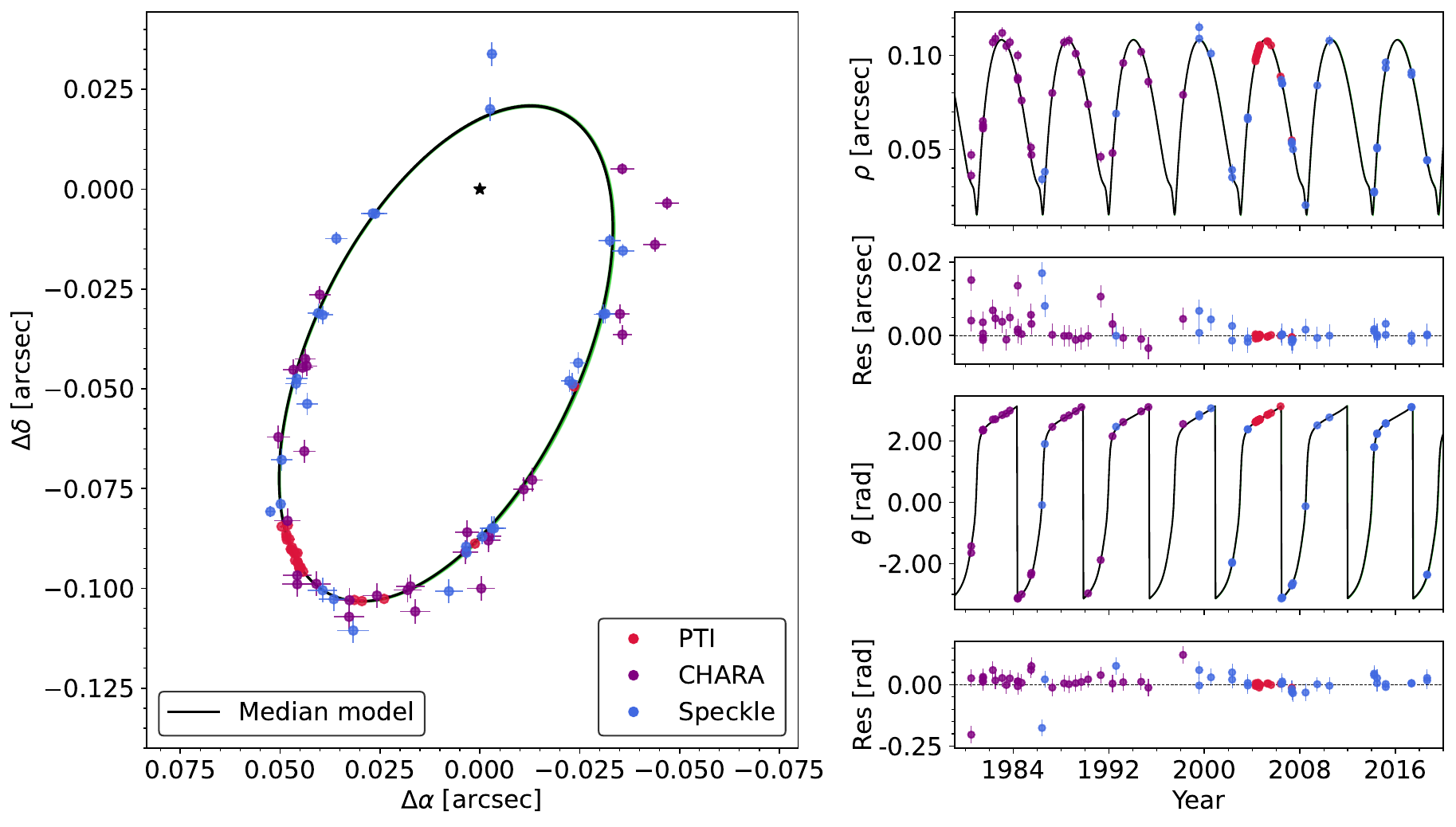}
    \caption{Relative \new{outer} orbit of V819 Her\new{, shown as the position of the inner binary (Ba+Bb) relative to the tertiary (A), following the WDS convention}. The left panel shows the orbit generated with the median parameters in black, while the green curves represent 1000 sample orbits randomly drawn from the posterior distribution. The star symbol at the origin marks the position of \new{the tertiary (A)}. The right panels present the angular separation and position angle as a function of time, together with the corresponding model fits and residuals.}
    \label{fig:astrometry_solution}
\end{figure*}

\subsection{Priors and Parameterisation}
\label{subsec:priors}

We adopted weakly informative priors centred on plausible scales for $a_{\rm ang}$, $\log P_3$, \new{the mass ratio ($q_3$)}, and the systemic velocity ($\gamma$). \new{A Beta prior \citep{kipping2013}, which is defined on $[0, 1]$ and favours low eccentricities, was used for the eccentricity ($e_3$), and a bounded normal distribution for} the observed parallax ($m_\pi$), while a uniform prior was applied to the cosine of the inclination ($\cos i_3$) to ensure an isotropic distribution. \new{To improve sampling efficiency,} the angular parameters ($\Omega_3, \omega_3$) were reparameterised as $p = (\Omega_3 + \omega_3)/2$ and $m = (\Omega_3 - \omega_3)/2$. The epoch of periastron passage ($T_{\rm per}$) was analogously sampled as \new{an orbital phase}.

To account for potential underestimation of the formal measurement errors and unmodelled instrumental or astrophysical scatter, \new{data set}-specific jitter terms were modelled in log-space and added in quadrature to the formal uncertainties of all observables. These additive noise terms ($\log \sigma_{\rho,{\rm jit}}$, $\log \sigma_{\theta,{\rm jit}}$, $\log \sigma_{{\rm RV},{\rm jit}}$, and $\log \sigma_{{\rm OC},{\rm jit}}$) were treated as free parameters and sampled simultaneously with the orbital solution.

\begin{revblock}
We do not adopt the inner-binary mass of \citet{muterspaugh2008}, $M_{12} = 2.560 \pm 0.067\,M_\odot$, as a constraint: it was derived from the same PHASES astrometry and the same radial velocities of \citet{scarfe1994} that enter our fit, so that including it would count these data twice. The effect of adding it is quantified in Section~\ref{subsec:sensitivity}.
\end{revblock}

Regarding the spatial scale, we adopted an informative prior on the observed Gaia DR3 parallax. Specifically, the parallax parameter was assigned a truncated normal prior centred at $13.815$ mas with $\sigma=0.496$ mas and bounds of $[0,50]$ mas. This prior reflects the Gaia DR3 catalogue uncertainty while allowing the joint astrometric, ETV, and RV model to infer \new{an orbital} parallax from the full orbital solution. The Gaia DR3 astrometric solution for V819\,Her has a high Renormalised Unit Weight Error (RUWE) of 11.85, indicating that the single-star solution is likely affected by unresolved orbital motion in this hierarchical system. The \new{orbital} parallax inferred by our coupled model is therefore constrained not only by the catalogue prior, but also by the physical consistency between the sky-plane orbit, the LiTE signal, and the tertiary radial velocities\new{; removing this prior changes $\pi_{\rm orb}$ by only $0.02$\,mas (Section~\ref{subsec:sensitivity})}.

Table \ref{tab:priors_joint_final_units} summarizes the prior distributions
adopted for all free parameters in the joint model.

\subsection{Inference Setup and Diagnostics}
\label{subsec:inference}

\begin{table}
\centering
\caption{Posterior orbital and physical parameters of the V819 Her tertiary component derived from the joint MCMC analysis. \new{The quoted uncertainties correspond to the 16th and 84th posterior percentiles.}}
\label{tab:results_table}
\setlength{\tabcolsep}{4pt}
\renewcommand{\arraystretch}{1.25}
\begin{revblock}
\begin{tabular}{lllc}
\toprule
Parameter & Symbol & Unit & Value \\
\midrule
\multicolumn{4}{c}{\textit{Orbital elements}} \\
\midrule
Orbital period & $P_3$ & d & $2016.97^{+0.26}_{-0.27}$ \\
Periastron epoch & $T_{\rm per}$ & BJD$_{\rm TDB}$ & $2452617.11^{+0.98}_{-0.96}$ \\
Eccentricity & $e_3$ & -- & $0.6800^{+0.0016}_{-0.0016}$ \\
Argument of periastron & $\omega_3$ & deg & $221.14^{+0.30}_{-0.30}$ \\
Longitude of asc. node & $\Omega_3$ & deg & $142.12^{+0.22}_{-0.23}$ \\
Inclination & $i_3$ & deg & $55.65^{+0.20}_{-0.20}$ \\
Angular semi-major axis & $a_{\rm ang}$ & mas & $73.57^{+0.12}_{-0.12}$ \\
Systemic velocity & $\gamma$ & km\,s$^{-1}$ & $-3.358^{+0.055}_{-0.055}$ \\
\midrule
\multicolumn{4}{c}{\textit{Physical properties}} \\
\midrule
Orbital parallax & $\pi_{\rm orb}$ & mas & $14.08^{+0.13}_{-0.13}$ \\
Mass ratio & $q_3$ & -- & $0.718^{+0.014}_{-0.014}$ \\
Total mass & $M_{\rm tot}$ & $M_\odot$ & $4.675^{+0.125}_{-0.119}$ \\
Inner-binary mass & $M_{12}$ & $M_\odot$ & $2.721^{+0.056}_{-0.054}$ \\
Tertiary mass & $M_3$ & $M_\odot$ & $1.954^{+0.072}_{-0.068}$ \\
\midrule
\multicolumn{4}{c}{\textit{Derived scales}} \\
\midrule
Relative semi-major axis & $a_{\rm rel}$ & AU & $5.224^{+0.046}_{-0.045}$ \\
Inner-binary semi-major axis & $a_{12}$ & AU & $2.184^{+0.042}_{-0.040}$ \\
Tertiary semi-major axis & $a_3$ & AU & $3.040^{+0.016}_{-0.016}$ \\
RV semi-amplitude & $K_3$ & km\,s$^{-1}$ & $18.464^{+0.091}_{-0.089}$ \\
LiTE amplitude & $A_{\rm LiTE}$ & d & $0.01041^{+0.00020}_{-0.00019}$ \\
\bottomrule
\end{tabular}
\end{revblock}
\end{table}

To sample from the joint posterior distribution, we employed the No-U-Turn Sampler \citep[NUTS;][]{hoffman2011}. The sampling was executed using version 5 of the \texttt{PyMC} probabilistic programming framework\footnote{\url{https://www.pymc.io}} \citep{pymc2023}.

For the final inference, we ran four independent Markov chains. Each chain used 3000 tuning steps, which were discarded, followed by 5000 posterior draws, yielding a total of 20000 posterior samples. \new{Throughout this paper, uncertainties correspond to the 16th and 84th posterior percentiles.}

\new{We assessed convergence with the Gelman--Rubin diagnostic \citep{gelman1992} and the effective sample size (ESS), and by visual inspection of the trace plots. No divergent transitions occurred, and all sampled parameters have $\widehat{R} \le 1.001$ and bulk and tail ESS above 19\,000 and 12\,000, respectively.}

The full posterior probability distributions and the correlations between the sampled orbital parameters are presented in Fig.~\ref{fig:corner_orbital} in Appendix \ref{sec:appendix_plots}, while the posteriors for the \new{orbital parallax, mass ratio and component masses} are provided in Fig.~\ref{fig:corner_physical}.

\subsection{Orbital Solution and Physical Properties}
\label{subsec:results}

\begin{figure*}
    \centering
    \includegraphics[width=\textwidth]{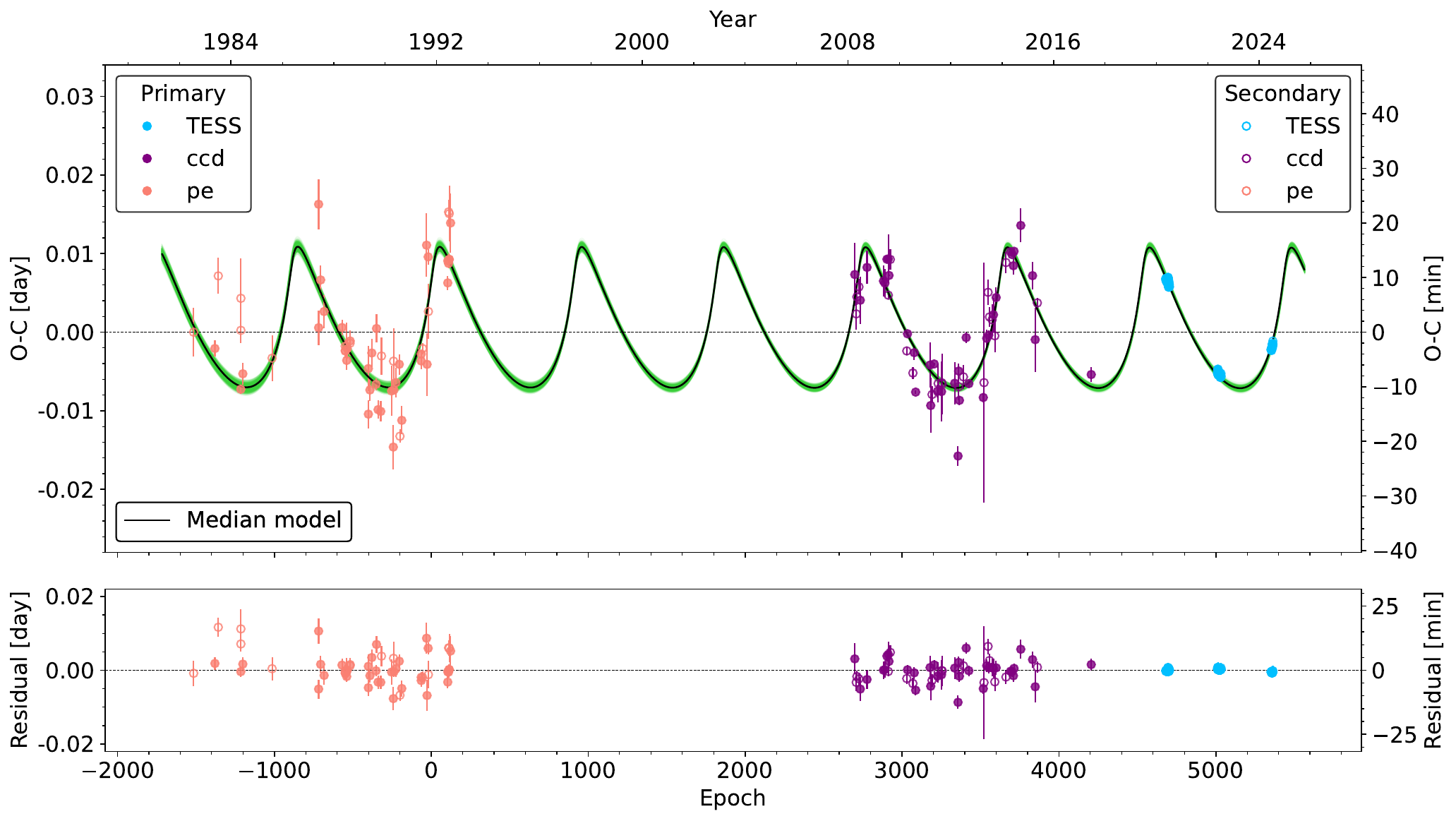}
    \caption{Eclipse timing variation solution for V819 Her. The upper panel shows the observed O$-$C values together with the median posterior model in black, while the green curves represent 1000 model realizations randomly drawn from the posterior distribution. The lower panel shows the residuals relative to the median model. Different colours indicate the timing data sources.}
    \label{fig:etv_solution}
\end{figure*}

\begin{figure*}
    \centering
    \includegraphics[width=\textwidth]{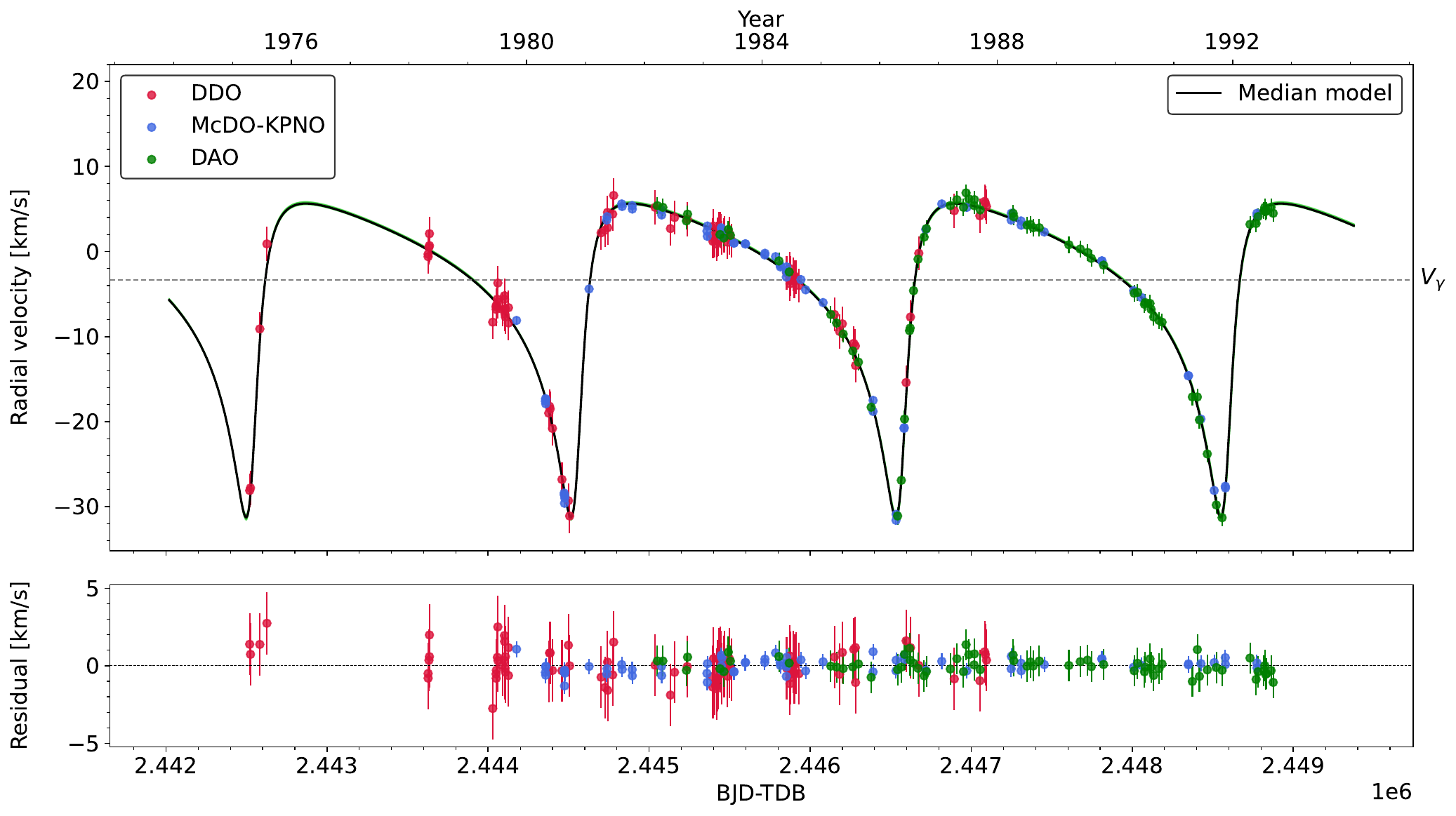}
    \caption{Radial velocity solution for the tertiary component of V819 Her. The upper panel shows the observed radial velocities together with the median posterior model in black, while the green curves represent 1000 model realizations randomly drawn from the posterior distribution. The lower panel shows the residuals relative to the median model. Different colours indicate the contributing observatories or instruments.}
    \label{fig:rv_solution}
\end{figure*}

\new{The joint analysis of the astrometric, ETV, and RV data constrains all orbital and physical parameters of the outer orbit.} The inferred median values and \new{their uncertainties} are presented in Table~\ref{tab:results_table}. The model fits to the three observational data sets are shown in Figs.~\ref{fig:astrometry_solution}, \ref{fig:etv_solution}, and \ref{fig:rv_solution}.

The posterior constraints from the three data sets are mutually consistent within the coupled orbital model, with the astrometric, ETV, and RV observables jointly determining the geometry and physical scale of the outer orbit. This consistency is illustrated by the astrometric, eclipse timing, and radial velocity fits shown in Figs. \ref{fig:astrometry_solution}, \ref{fig:etv_solution}, and \ref{fig:rv_solution}.

\begin{revblock}
\subsection{Prior Sensitivity and Consistency Checks}
\label{subsec:sensitivity}

To verify that our conclusions are driven by the data rather than by the priors, we repeated the inference with modified assumptions, keeping all other settings identical (Table~\ref{tab:prior_sensitivity}). Replacing the Beta eccentricity prior with a uniform prior, Beta$(1, 1)$, or with the truncated normal prior of the submitted analysis, $\mathcal{TN}(0.7, 0.2)$ on $[0, 1]$, leaves all parameters unchanged ($e_3 = 0.6801 \pm 0.0016$ in both cases), although the three priors assign 5, 40 and 67 per cent of their probability, respectively, to $e > 0.6$. Replacing the Gaia DR3 parallax prior with a uniform prior on $[1, 50]$\,mas gives $\pi_{\rm orb} = 14.10 \pm 0.13$\,mas, $M_{12} = 2.713 \pm 0.056\,M_\odot$ and $M_3 = 1.945 \pm 0.072\,M_\odot$, so that the orbital parallax is determined by the data and its comparison with the Gaia catalogues (Section~\ref{sec:discussion}) is not circular. Adding the inner-binary mass of \citet{muterspaugh2008} as an external measurement shifts $M_{12}$, $M_3$ and $\pi_{\rm orb}$ by $-0.06\,M_\odot$, $-0.07\,M_\odot$ and $+0.14$\,mas, respectively, all within the uncertainties of the fiducial solution (Table~\ref{tab:results_table}).

Further checks use information that does not enter the fit or subsets of the data. First, the angular semi-major axis of the inner Ba--Bb orbit predicted by Kepler's third law, $0.657 \pm 0.002$\,mas, agrees ($0.3\sigma$) with the value of $0.663 \pm 0.022$\,mas measured independently with the CHARA Array by \citet{obrien2011}. As the parallax cancels in this prediction, which depends only on $a_{\rm ang}$, $P_3$ and $q_3$, it confirms the consistency of the outer-orbit geometry but does not test the absolute masses. Second, we fitted the astrometry alone, replacing the PTI positions with the PHASES measurements in $(\Delta\alpha, \Delta\delta)$ with their full error ellipses \citep{muterspaugh2010}. This fit gives $P_3 = 2016.1 \pm 0.7$\,d, $T_{\rm per} = 2452617.5 \pm 2.2$ (BJD), $e_3 = 0.6767 \pm 0.0010$, $\omega_3 = 220.7 \pm 0.3\degr$, $\Omega_3 = 142.5 \pm 0.2\degr$ and $i_3 = 55.7 \pm 0.2\degr$, consistent with the joint solution; the eccentricity differs by $1.8\sigma$.
\begin{table}
\begin{revblock}
\centering
\caption{Prior sensitivity of the joint solution. All runs use the same data and sampler settings.}
\label{tab:prior_sensitivity}
\setlength{\tabcolsep}{3pt}
\renewcommand{\arraystretch}{1.25}
\footnotesize
\begin{tabular*}{\columnwidth}{@{}l@{\extracolsep{\fill}}cccc@{}}
\toprule
Run & $e_3$ & $\pi_{\rm orb}$ [mas] & $M_{12}$ [$M_\odot$] & $M_3$ [$M_\odot$] \\
\midrule
Fiducial & $0.6800^{+0.0016}_{-0.0016}$ & $14.08^{+0.13}_{-0.13}$ & $2.721^{+0.056}_{-0.054}$ & $1.954^{+0.072}_{-0.068}$ \\
$e_3$: Beta$(1,1)$ & $0.6801^{+0.0016}_{-0.0016}$ & $14.08^{+0.13}_{-0.13}$ & $2.721^{+0.055}_{-0.054}$ & $1.956^{+0.070}_{-0.069}$ \\
$e_3$: $\mathcal{TN}$ & $0.6801^{+0.0016}_{-0.0016}$ & $14.08^{+0.13}_{-0.13}$ & $2.721^{+0.056}_{-0.054}$ & $1.955^{+0.071}_{-0.068}$ \\
$\pi$: flat & $0.6800^{+0.0016}_{-0.0016}$ & $14.10^{+0.13}_{-0.13}$ & $2.713^{+0.057}_{-0.055}$ & $1.945^{+0.073}_{-0.070}$ \\
$M_{12}$: M08 & $0.6800^{+0.0016}_{-0.0016}$ & $14.23^{+0.10}_{-0.10}$ & $2.659^{+0.041}_{-0.042}$ & $1.883^{+0.056}_{-0.055}$ \\
\bottomrule
\end{tabular*}
\par\vspace{6pt}
\parbox{\columnwidth}{\footnotesize Fiducial: Beta$(0.867, 3.03)$ prior on $e_3$ and Gaia DR3 prior on $\pi$. $e_3$: Beta$(1,1)$ and $e_3$: $\mathcal{TN}$: uniform and truncated normal $\mathcal{TN}(0.7, 0.2)$ on $[0,1]$ priors on $e_3$. $\pi$: flat: uniform prior on $[1, 50]$\,mas instead of the Gaia DR3 prior. $M_{12}$: M08: $M_{12} = 2.560 \pm 0.067\,M_\odot$ of \citet{muterspaugh2008} included as an external measurement.}
\end{revblock}
\end{table}
 A fit to the ETV and RV data alone gives $P_3 = 2018.4 \pm 0.7$\,d and $T_{\rm per} = 2452618.1 \pm 2.4$, so that the sky-plane and line-of-sight data determine the same periastron epoch independently; their periods differ by $2.3\sigma$, and the joint value lies between them. Finally, we refitted the data including the 187 radial velocities of the brighter inner-binary component Ba from \citet{scarfe1994}, modelled as the sum of a circular inner orbit and the outer-orbit motion of the inner binary. The solution is unchanged ($\pi_{\rm orb} = 14.08 \pm 0.12$\,mas, $M_{12} = 2.721 \pm 0.053\,M_\odot$, $M_3 = 1.955 \pm 0.066\,M_\odot$), as these velocities carry little weight because of their large scatter, which was already noted by \citet{scarfe1994}.

\end{revblock}

\section{Discussion and Conclusions}
\label{sec:discussion}

The joint astrometry--ETV--RV model provides a dynamically coupled solution for the outer orbit of V819\,Her, linking the sky-plane orbit, the LiTE signal, and the tertiary radial velocities through a common Keplerian geometry. This coupling allows the angular orbit, the line-of-sight motion, and the physical velocity scale to be interpreted within the same dynamical framework, rather than as separate orbital solutions.

The availability of all three constraints for V819\,Her is unusual: the resolved astrometric orbit fixes the sky-plane geometry, the ETV signal probes the light-time displacement of the inner binary, and the tertiary RVs provide the line-of-sight velocity scale. Their joint use is therefore what allows the \new{orbital} parallax and mass partitioning to be constrained self-consistently.

\begin{revblock}
An important outcome of this joint model is the orbital parallax, $\pi_{\rm orb} = 14.08 \pm 0.13$\,mas, which follows geometrically from the angular orbit and the physical orbit sizes set by the RV and LiTE amplitudes, without any assumption on the masses. The Gaia DR3 catalogue reports a raw parallax of $\pi_{\rm DR3} = 13.815 \pm 0.496$\,mas with a high RUWE of 11.85. V819\,Her lacks a Non-Single Star (NSS) solution in Gaia DR3 (\texttt{non\_single\_star = 0}), so that no orbital or acceleration model was fitted to account for the photocentric motion. As discussed by \citet{groenewegen2023}, the limited availability of NSS solutions in DR3 for known binary and multiple systems often leaves a perturbed single-star solution as the only Gaia DR3 parallax. V819\,Her is part of the sample of \citet{groenewegen2023}, who derived an orbital parallax of $14.584 \pm 0.201$\,mas from the visual orbit of \citet{muterspaugh2008} and the spectroscopic amplitudes of \citet{scarfe1994}. Applying the zero-point correction of \citet{lindegren2021} yields $\pi_{\rm DR3,corr} = 13.840 \pm 0.496$\,mas, consistent with our orbital parallax ($0.5\sigma$, also for the fit without the Gaia prior; Section~\ref{subsec:sensitivity}); a full astrometric-orbital solution will have to await Gaia DR4.

For comparison, the earlier single-star solutions of Gaia DR2 ($\pi_{\rm DR2} = 14.132 \pm 0.302$\,mas; \citealt{gaiadr2}) and the HIPPARCOS New Reduction \citep[$13.52 \pm 0.93$\,mas;][]{vanleeuwen2007} also agree with our value within $1\sigma$. The DR2 parallaxes are subject to a global zero-point offset of approximately $-0.029$~mas as determined from quasars by \citet{lindegren2018}; independent studies using stellar populations of comparable brightness to V819\,Her find offsets in the range $-0.05$ to $-0.08$~mas \citep{stassun2018, zinn2019}, a systematic uncertainty that is small relative to the DR2 formal error. Our orbital parallax is smaller than the value of \citet{groenewegen2023} and those of \citet{scarfe1994} and \citet{muterspaugh2008} ($14.5 \pm 0.2$ and $14.57 \pm 0.19$\,mas) by about $2\sigma$. The difference arises from our updated angular orbit (smaller $a_{\rm ang}$ and $i_3$) and from the size of the inner binary's barycentric orbit, which our model takes from the LiTE amplitude rather than from the spectroscopic amplitude of its centre of mass \citep{scarfe1994}. The eclipse timings correspond to $K_{\rm B} = 13.26 \pm 0.24$\,km\,s$^{-1}$, consistent within $1\sigma$ with the spectroscopic value of $12.89 \pm 0.32$\,km\,s$^{-1}$. The larger value is already preferred by the pre-TESS timings alone ($13.8 \pm 0.6$\,km\,s$^{-1}$); the TESS minima mainly reduce its uncertainty. Including the Ba velocities in the fit leaves the solution unchanged (Section~\ref{subsec:sensitivity}).

The Gaia DR3 spectroscopic pipeline reports a median radial velocity of $2.56 \pm 1.28$\,km\,s$^{-1}$ and a total range of $\texttt{rv\_amplitude\_robust} = 14.7$\,km\,s$^{-1}$ over 20 RVS epochs, obtained with a single template ($T_{\rm eff} = 5500$\,K, $\log g = 3.5$) that matches the evolved component A. Over the Gaia DR3 observing window (2014.56--2017.40) the tertiary is near apastron, and our orbit predicts a median velocity of $3.5$\,km\,s$^{-1}$, consistent with the Gaia value, and a total velocity change of only $\approx 7$\,km\,s$^{-1}$. The measured range is therefore far smaller than the $>150$\,km\,s$^{-1}$ excursions of the inner binary, which are largely smeared out in a single-template cross-correlation, but about twice the variation expected from the outer orbit alone. We attribute the excess to partial blending of the inner-binary lines with those of component A.
\end{revblock}

\new{The derived orbital elements are compared with previous determinations in Table~\ref{tab:literature_comparison}. The joint fit determines the outer period, the periastron epoch, the orbital parallax and both component masses more precisely than any of the earlier solutions, and the masses are the first to combine the sky-plane, light-time and radial-velocity constraints, while the eccentricity and the node agree with the PHASES-based solution of \citet{muterspaugh2008} within $1\sigma$.} We find a high eccentricity of \new{$e_3 = 0.680$} for the 5.5-year outer orbit, \new{confirming the high eccentricity first found by} \citet{scarfe1994}. The derived LiTE amplitude of $A_{\mathrm{LiTE}} \approx 0.010$ days and the velocity semi-amplitude \new{$K_3 \approx 18.5$} km s$^{-1}$ are self-consistently tied to the component masses.

\begin{revblock}
A central result of this work is that the component masses follow from the joint fit without external mass information: $M_{12} = 2.721 \pm 0.055\,M_\odot$ and $M_3 = 1.954 \pm 0.070\,M_\odot$. They agree with those of \citet{scarfe1994} within $1\sigma$ and exceed those of \citet{muterspaugh2008} by less than $2\sigma$, because our physical orbit is larger (Table~\ref{tab:literature_comparison}). Relative to \citet{muterspaugh2008}, about 60 per cent of the difference in $M_3$ comes from the inner-binary orbit set by the eclipse timings, which enters both $a_{\rm rel}$ and $q_3$, and about 30 per cent from our slightly lower inclination ($55.65\degr$ versus $56.40\degr$); the same larger orbit yields our smaller orbital parallax. The inclination difference is tied to the periastron epoch. The period of \citet{muterspaugh2008}, $2019.66 \pm 0.35$\,d, is longer than ours by $6\sigma$ and places the 2002 periastron about 11\,d later ($7\sigma$; Table~\ref{tab:literature_comparison}). As the PHASES data cover only 2004--2007, $T_{\rm per}$ is correlated with $i_3$ and $a_{\rm ang}$ over this short arc. Our fits to the astrometry alone and to the ETV and RV data alone give the same, earlier periastron epoch independently (Section~\ref{subsec:sensitivity}), while the 1980--2024 baseline of the speckle and timing data fixes $P_3$. The eccentricity and node of \citet{scarfe1994} differ from ours by $3\sigma$ and $4\sigma$, whereas those of \citet{muterspaugh2008} agree within $1\sigma$. The mass of the tertiary component is higher than expected for a standard G8-type main-sequence star, suggesting that the third body is in a more advanced evolutionary stage.

Recently, \citet{howell2024} presented Gemini North/`Alopeke speckle observations of V819\,Her at 832\,nm reduced with Multi-Frame Blind Deconvolution (MFBD). Their reconstructed image (their fig.~4) resolves the Ba+Bb pair from component A at $\Delta\alpha \approx +40$\,mas, $\Delta\delta \approx -100$\,mas ($\rho \approx 108$\,mas, $\theta \approx 158\degr$). As the observing epoch and calibrated astrometry are not reported, we do not include this position in the fit. Our orbit reproduces this configuration, within the $\sim$10\,mas pixel scale of the image, only between 2021.1 and 2021.9, close to the 2021.7 apastron; at 2021.5 we predict $\rho = 107.9$\,mas and $\theta = 158.6\degr$. \citet{howell2024} also report a fainter source at 102\,mas to the north-west, close to $180\degr$ in position angle from Ba+Bb; they note this coincidence but argue that the source is not a reconstruction ghost, since the MFBD technique does not produce the $180\degr$ ambiguity of standard speckle interferometry. As it rests on a single epoch, multi-epoch high-resolution imaging or Gaia DR4 epoch astrometry will be needed to test whether it is a bound fourth component.
\end{revblock}

\begin{revblock}
\begin{table*}
\begin{revblock}
\centering
\caption{Outer-orbit and physical parameters of V819\,Her from this work compared with previous determinations. Uncertainties of this work are the 16th--84th posterior percentiles; literature values are quoted with their published uncertainties.}
\label{tab:literature_comparison}
\setlength{\tabcolsep}{6pt}
\renewcommand{\arraystretch}{1.15}
\small
\begin{tabular}{llcccc}
\toprule
Parameter & Unit & This work & Zasche et al. (2014) & Muterspaugh et al. (2008) & Scarfe et al. (1994) \\
\midrule
$P_3$ & d & $2016.97 \pm 0.27$ & $2015.8 \pm 54.9$ & $2019.66 \pm 0.35$ & $2018.8 \pm 0.7$ \\
$T_{\rm per}$ & BJD$-2400000$ & $52617.11 \pm 0.97$ & $52621.6 \pm 81.2^{a}$ & $52628.0 \pm 1.3^{a}$ & $46564.6 \pm 1.0^{a,b}$ \\
$e_3$ & -- & $0.6800 \pm 0.0016$ & $0.687 \pm 0.011$ & $0.67974 \pm 0.00066$ & $0.672 \pm 0.002$ \\
$\omega_3$ & deg & $221.14 \pm 0.30$ & $223.8 \pm 3.5$ & $222.50 \pm 0.22$ & $220.6 \pm 0.3^{c}$ \\
$\Omega_3$ & deg & $142.12 \pm 0.22$ & $140.8 \pm 5.9$ & $141.96 \pm 0.12$ & $143.7 \pm 0.3^{c}$ \\
$i_3$ & deg & $55.65 \pm 0.20$ & $56.82 \pm 1.63$ & $56.40 \pm 0.13$ & $56.2 \pm 0.4$ \\
$a_{\rm ang}$ & mas & $73.57 \pm 0.12$ & $74.6 \pm 3.7$ & $74.41 \pm 0.67^{d}$ & $75.0 \pm 0.5$ \\
$a_{\rm rel}$ & AU & $5.224 \pm 0.045$ & $5.13 \pm 0.25^{e}$ & $5.108 \pm 0.046$ & $5.16 \pm 0.06$ \\
$\pi_{\rm orb}$ & mas & $14.08 \pm 0.13$ & $14.53 \pm 0.38^{f}$ & $14.57 \pm 0.19$ & $14.5 \pm 0.2$ \\
$K_3$ & km\,s$^{-1}$ & $18.464 \pm 0.090$ & -- & -- & $18.32 \pm 0.07$ \\
$A_{\rm LiTE}$ & d & $0.01041 \pm 0.00019$ & $0.0090 \pm 0.0008$ & -- & -- \\
$\gamma$ & km\,s$^{-1}$ & $-3.358 \pm 0.055$ & -- & $-3.375 \pm 0.059^{g}$ & $-3.37 \pm 0.04$ \\
$M_{12}$ & $M_\odot$ & $2.721 \pm 0.055$ & -- & $2.560 \pm 0.067$ & $2.64 \pm 0.08^{h}$ \\
$M_3$ & $M_\odot$ & $1.954 \pm 0.070$ & $1.86 \pm 0.30$ & $1.799 \pm 0.098$ & $1.86 \pm 0.09$ \\
\bottomrule
\end{tabular}
\vspace{3pt}
\begin{flushleft}
\footnotesize
$^{a}$ Converted to BJD$_{\rm TDB}$: \citet{scarfe1994} and \citet{zasche2014} give HJD, and \citet{muterspaugh2008} give MHJD $= 52627.5$. The correction from HJD to BJD$_{\rm TDB}$ ($\lesssim 0.001$\,d) is negligible at the quoted precision. \\
$^{b}$ Periastron passage of 1986, three orbital cycles before the other entries. \\
$^{c}$ Published as $\omega = 40.6^\circ$ for the centre of mass of the close pair and $\Omega = 323.7^\circ$ for the ascending node of the brighter component; both angles are shifted by $180^\circ$ to the $(\omega_3, \Omega_3)$ convention used here, which leaves the sky-projected orbit unchanged. \\
$^{d}$ As listed for this orbit by \citet{groenewegen2023}. \\
$^{e}$ Sum of the published barycentric semi-major axes $a_{12} = 2.16 \pm 0.15$ and $a_3 = 2.97 \pm 0.20$\,AU. \\
$^{f}$ Inverse of the published distance. \\
$^{g}$ Value for the McDonald/Kitt Peak data set; the DAO and DDO values are $-3.385 \pm 0.065$ and $-3.35 \pm 0.12$\,km\,s$^{-1}$. \\
$^{h}$ Sum of the component masses $M_{\rm Ba} = 1.53 \pm 0.08$ and $M_{\rm Bb} = 1.11 \pm 0.02\,M_\odot$.
\end{flushleft}
\end{revblock}
\end{table*}
\end{revblock}

Previous interferometric studies have constrained the mutual inclination between the inner and outer orbits. The present model does not rederive this quantity, because the orientation of the inner eclipsing binary orbit is not fitted here. Our analysis instead focuses on the outer orbit and on its dynamical coupling to the ETV and RV observables. This joint modelling approach demonstrates the value of combining complementary observational techniques when characterising hierarchical systems for which single-catalogue astrometric solutions may be affected by unresolved orbital motion. Finally, we provide predicted relative astrometric positions of \new{the inner binary with respect to the tertiary, and their uncertainties,} for five years beyond 2026.5 in Table~\ref{tab:future_astrometry}.

\begin{table}
    \centering
    \caption{Predicted relative astrometric positions of \new{the inner binary (Ba+Bb) with respect to the tertiary (A)} in V819 Her\new{, following the WDS convention,} for the five-year interval following 2026.5\new{, given as the median and the 16th--84th percentile range of 4000 posterior draws}.}
    \label{tab:future_astrometry}
    \renewcommand{\arraystretch}{1.25}
    \begin{revblock}
    \begin{tabular}{ccc}
        \toprule
        Epoch [yr] & $\rho$ [mas] & $\theta$ [deg] \\
        \midrule
        2027.0 & $107.74^{+0.07}_{-0.07}$ & $158.35^{+0.06}_{-0.06}$ \\
        2027.5 & $107.03^{+0.10}_{-0.10}$ & $164.61^{+0.06}_{-0.07}$ \\
        2028.0 & $100.14^{+0.13}_{-0.14}$ & $171.35^{+0.08}_{-0.09}$ \\
        2028.5 & $87.89^{+0.17}_{-0.17}$  & $179.59^{+0.11}_{-0.12}$ \\
        2029.0 & $71.11^{+0.21}_{-0.21}$  & $191.19^{+0.16}_{-0.17}$ \\
        2029.5 & $51.46^{+0.25}_{-0.23}$  & $211.05^{+0.28}_{-0.27}$ \\
        2030.0 & $34.77^{+0.19}_{-0.20}$  & $253.32^{+0.61}_{-0.60}$ \\
        2030.5 & $23.25^{+0.40}_{-0.43}$  & $333.63^{+1.57}_{-1.54}$ \\
        2031.0 & $48.21^{+0.50}_{-0.50}$  & $125.49^{+0.32}_{-0.33}$ \\
        2031.5 & $82.61^{+0.24}_{-0.25}$  & $142.79^{+0.11}_{-0.11}$ \\
        \bottomrule
    \end{tabular}
    \end{revblock}
\end{table}

\new{In conclusion, the joint analysis yields a self-consistent high-eccentricity outer orbit with $P_3 = 2016.97$\,d and $e_3 = 0.680$, an orbital parallax of $\pi_{\rm orb} = 14.08$\,mas, and dynamical masses of $M_{12} = 2.72\,M_\odot$ for the inner binary and $M_3 = 1.95\,M_\odot$ for the tertiary.} These results are consistent with independent astrometric catalogues\new{, although the Gaia DR3 single-star solution is likely affected by unresolved orbital motion}. The relatively high tertiary mass compared with a standard G8 main-sequence classification suggests that the tertiary component may be in a more evolved state. Continued astrometric and spectroscopic monitoring will be valuable for testing the predicted positions listed in Table~\ref{tab:future_astrometry} and for further refining the dynamical architecture of the system. \new{In particular, high-resolution astrometry around the periastron passages, of which the two most recent (2019.5 and 2025.0) are not covered by the present astrometric data and the next occurs at 2030.5, together with new eclipse timings and the Gaia DR4 epoch astrometry and radial velocities, will sharpen the outer period and the orbital parallax and hence the dynamical masses.}

\section*{Acknowledgements}

We gratefully acknowledge the support by the Y\"OK ADEP project FBA-2023-2698. This research has made use of the Washington Double Star Catalog maintained at the U.S. Naval Observatory. This work has made use of data from the European Space Agency (ESA) mission Gaia, processed by the Gaia Data Processing and Analysis Consortium (DPAC). Funding for the DPAC has been provided by national institutions, in particular the institutions participating in the Gaia Multilateral Agreement. This research has made use of the HIPPARCOS catalogue (ESA 1997) and the HIPPARCOS New Reduction \citep{vanleeuwen2007} accessible via the VizieR catalogue access tool, CDS, Strasbourg, France. This paper includes data collected by the TESS mission, which are publicly available from the Mikulski Archive for Space Telescopes (MAST). Funding for the TESS mission is provided by NASA's Science Mission Directorate.

\section*{Data Availability}
\label{subsec:data_availability}
The data underlying this article are available from both public archives and as supplementary online material. The public observational data sets analysed in this study were obtained from the European Space Agency (ESA) Gaia Archive\footnote{https://gea.esac.esa.int/archive/}, the HIPPARCOS New Reduction catalogue via the VizieR Service\footnote{https://vizier.cds.unistra.fr/viz-bin/VizieR-3?-source=I/311}, \new{the PHASES differential astrometry archive via the VizieR Service (J/AJ/140/1579),} the TESS archive via the Mikulski Archive for Space Telescopes (MAST)\footnote{https://mast.stsci.edu/portal/Mashup/Clients/Mast/Portal.html}, and the Washington Double Star (WDS)\footnote{https://www.astro.gsu.edu/wds/} Catalog maintained by the U.S. Naval Observatory. The astrometric measurements, the times of minima (including the new TESS timings) and the radial velocities used in this study are available as supplementary material to this article and will also be made available at the CDS.

\bibliographystyle{mnras}
\bibliography{v819her}

@ARTICLE{astroquery2019,
   author = {{Ginsburg}, A. and {Sip{\H o}cz}, B.~M. and {Brasseur}, C.~E. and
	{Cowperthwaite}, P.~S. and {Craig}, M.~W. and {Deil}, C. and
	{Guillochon}, J. and {Guzman}, G. and {Liedtke}, S. and {Lian Lim}, P. and
	{Lockhart}, K.~E. and {Mommert}, M. and {Morris}, B.~M. and
	{Norman}, H. and {Parikh}, M. and {Persson}, M.~V. and {Robitaille}, T.~P. and
	{Segovia}, J.-C. and {Singer}, L.~P. and {Tollerud}, E.~J. and
	{de Val-Borro}, M. and {Valtchanov}, I. and {Woillez}, J. and
	{The Astroquery collaboration} and {a subset of the astropy collaboration}
	},
    title = "{astroquery: An Astronomical Web-querying Package in Python}",
  journal = {\aj},
archivePrefix = "arXiv",
   eprint = {1901.04520},
 primaryClass = "astro-ph.IM",
     year = 2019,
    month = mar,
   volume = 157,
      eid = {98},
    pages = {98},
      doi = {10.3847/1538-3881/aafc33},
   adsurl = {https://adsabs.harvard.edu/abs/2019AJ....157...98G}
}

@ARTICLE{boyd1985,
       author = {{Boyd}, L.~J. and {Genet}, R.~M. and {Hall}, D.~S. and {Persinger}, W.~T. and {Fried}, R.~E. and {Wasson}, N.~F. and {Stelzer}, H.~J. and {Lines}, R.~D. and {Brooks}, P.~A. and {Hoff}, D.},
        title = "{Discovery of Eclipses and Long-Period Variability in the Triple System HR 6469}",
      journal = {Information Bulletin on Variable Stars},
         year = 1985,
        month = feb,
       volume = {2675},
        pages = {1},
       adsurl = {https://ui.adsabs.harvard.edu/abs/1985IBVS.2675....1B}
}

@ARTICLE{gaiadr2,
       author = {{Gaia Collaboration} and {Brown}, A.~G.~A. and {Vallenari}, A. and {Prusti}, T. and {de Bruijne}, J.~H.~J. and {Babusiaux}, C. and {Bailer-Jones}, C.~A.~L. and {Biermann}, M. and {Evans}, D.~W. and {Eyer}, L. and {Jansen}, F. and {Jordi}, C. and {Klioner}, S.~A. and {Lammers}, U. and {Lindegren}, L. and {Luri}, X. and {Mignard}, F. and {Panem}, C. and {Pourbaix}, D. and {Randich}, S. and {Sartoretti}, P. and {Siddiqui}, H.~I. and {Soubiran}, C. and {van Leeuwen}, F. and {Walton}, N.~A. and {Arenou}, F. and {Bastian}, U. and {Cropper}, M. and {Drimmel}, R. and {Katz}, D. and {Lattanzi}, M.~G. and {Bakker}, J. and {Cacciari}, C. and {Casta{\~n}eda}, J. and {Chaoul}, L. and {Cheek}, N. and {De Angeli}, F. and {Fabricius}, C. and {Guerra}, R. and {Holl}, B. and {Masana}, E. and {Messineo}, R. and {Mowlavi}, N. and {Nienartowicz}, K. and {Panuzzo}, P. and {Portell}, J. and {Riello}, M. and {Seabroke}, G.~M. and {Tanga}, P. and {Th{\'e}venin}, F. and {Gracia-Abril}, G. and {Comoretto}, G. and {Garcia-Reinaldos}, M. and {Teyssier}, D. and {Altmann}, M. and {Andrae}, R. and {Audard}, M. and {Bellas-Velidis}, I. and {Benson}, K. and {Berthier}, J. and {Blomme}, R. and {Burgess}, P. and {Busso}, G. and {Carry}, B. and {Cellino}, A. and {Clementini}, G. and {Clotet}, M. and {Creevey}, O. and {Davidson}, M. and {De Ridder}, J. and {Delchambre}, L. and {Dell'Oro}, A. and {Ducourant}, C. and {Fern{\'a}ndez-Hern{\'a}ndez}, J. and {Fouesneau}, M. and {Fr{\'e}mat}, Y. and {Galluccio}, L. and {Garc{\'\i}a-Torres}, M. and {Gonz{\'a}lez-N{\'u}{\~n}ez}, J. and {Gonz{\'a}lez-Vidal}, J.~J. and {Gosset}, E. and {Guy}, L.~P. and {Halbwachs}, J.-L. and {Hambly}, N.~C. and {Harrison}, D.~L. and {Hern{\'a}ndez}, J. and {Hestroffer}, D. and {Hodgkin}, S.~T. and {Hutton}, A. and {Jasniewicz}, G. and {Jean-Antoine-Piccolo}, A. and {Jordan}, S. and {Korn}, A.~J. and {Krone-Martins}, A. and {Lanzafame}, A.~C. and {Lebzelter}, T. and {L{\"o}ffler}, W. and {Manteiga}, M. and {Marrese}, P.~M. and {Mart{\'\i}n-Fleitas}, J.~M. and {Moitinho}, A. and {Mora}, A. and {Muinonen}, K. and {Osinde}, J. and {Pancino}, E. and {Pauwels}, T. and {Petit}, J.-M. and {Recio-Blanco}, A. and {Richards}, P.~J. and {Rimoldini}, L. and {Robin}, A.~C. and {Sarro}, L.~M. and {Siopis}, C. and {Smith}, M. and {Sozzetti}, A. and {S{\"u}veges}, M. and {Torra}, J. and {van Reeven}, W. and {Abbas}, U. and {Abreu Aramburu}, A. and {Accart}, S. and {Aerts}, C. and {Altavilla}, G. and {{\'A}lvarez}, M.~A. and {Alvarez}, R. and {Alves}, J. and {Anderson}, R.~I. and {Andrei}, A.~H. and {Anglada Varela}, E. and {Antiche}, E. and {Antoja}, T. and {Arcay}, B. and {Astraatmadja}, T.~L. and {Bach}, N. and {Baker}, S.~G. and {Balaguer-N{\'u}{\~n}ez}, L. and {Balm}, P. and {Barache}, C. and {Barata}, C. and {Barbato}, D. and {Barblan}, F. and {Barklem}, P.~S. and {Barrado}, D. and {Barros}, M. and {Barstow}, M.~A. and {Bartholom{\'e} Mu{\~n}oz}, S. and {Bassilana}, J.-L. and {Becciani}, U. and {Bellazzini}, M. and {Berihuete}, A. and {Bertone}, S. and {Bianchi}, L. and {Bienaym{\'e}}, O. and {Blanco-Cuaresma}, S. and {Boch}, T. and {Boeche}, C. and {Bombrun}, A. and {Borrachero}, R. and {Bossini}, D. and {Bouquillon}, S. and {Bourda}, G. and {Bragaglia}, A. and {Bramante}, L. and {Breddels}, M.~A. and {Bressan}, A. and {Brouillet}, N. and {Br{\"u}semeister}, T. and {Brugaletta}, E. and {Bucciarelli}, B. and {Burlacu}, A. and {Busonero}, D. and {Butkevich}, A.~G. and {Buzzi}, R. and {Caffau}, E. and {Cancelliere}, R. and {Cannizzaro}, G. and {Cantat-Gaudin}, T. and {Carballo}, R. and {Carlucci}, T. and {Carrasco}, J.~M. and {Casamiquela}, L. and {Castellani}, M. and {Castro-Ginard}, A. and {Charlot}, P. and {Chemin}, L. and {Chiavassa}, A. and {Cocozza}, G. and {Costigan}, G. and {Cowell}, S. and {Crifo}, F. and {Crosta}, M. and {Crowley}, C. and {Cuypers}, J. and {Dafonte}, C. and {Damerdji}, Y. and {Dapergolas}, A. and {David}, P. and {David}, M. and {de Laverny}, P. and {De Luise}, F.},
        title = "{Gaia Data Release 2. Summary of the contents and survey properties}",
      journal = {\aap},
         year = 2018,
        month = aug,
       volume = {616},
          eid = {A1},
        pages = {A1},
          doi = {10.1051/0004-6361/201833051},
archivePrefix = {arXiv},
       eprint = {1804.09365},
 primaryClass = {astro-ph.GA},
       adsurl = {https://ui.adsabs.harvard.edu/abs/2018A&A...616A...1G}
}

@ARTICLE{gaiadr3,
       author = {{Gaia Collaboration} and {Vallenari}, A. and {Brown}, A.~G.~A. and {Prusti}, T. and {de Bruijne}, J.~H.~J. and {Arenou}, F. and {Babusiaux}, C. and {Biermann}, M. and {Creevey}, O.~L. and {Ducourant}, C. and {Evans}, D.~W. and {Eyer}, L. and {Guerra}, R. and {Hutton}, A. and {Jordi}, C. and {Klioner}, S.~A. and {Lammers}, U.~L. and {Lindegren}, L. and {Luri}, X. and {Mignard}, F. and {Panem}, C. and {Pourbaix}, D. and {Randich}, S. and {Sartoretti}, P. and {Soubiran}, C. and {Tanga}, P. and {Walton}, N.~A. and {Bailer-Jones}, C.~A.~L. and {Bastian}, U. and {Drimmel}, R. and {Jansen}, F. and {Katz}, D. and {Lattanzi}, M.~G. and {van Leeuwen}, F. and {Bakker}, J. and {Cacciari}, C. and {Casta{\~n}eda}, J. and {De Angeli}, F. and {Fabricius}, C. and {Fouesneau}, M. and {Fr{\'e}mat}, Y. and {Galluccio}, L. and {Guerrier}, A. and {Heiter}, U. and {Masana}, E. and {Messineo}, R. and {Mowlavi}, N. and {Nicolas}, C. and {Nienartowicz}, K. and {Pailler}, F. and {Panuzzo}, P. and {Riclet}, F. and {Roux}, W. and {Seabroke}, G.~M. and {Sordo}, R. and {Th{\'e}venin}, F. and {Gracia-Abril}, G. and {Portell}, J. and {Teyssier}, D. and {Altmann}, M. and {Andrae}, R. and {Audard}, M. and {Bellas-Velidis}, I. and {Benson}, K. and {Berthier}, J. and {Blomme}, R. and {Burgess}, P.~W. and {Busonero}, D. and {Busso}, G. and {C{\'a}novas}, H. and {Carry}, B. and {Cellino}, A. and {Cheek}, N. and {Clementini}, G. and {Damerdji}, Y. and {Davidson}, M. and {de Teodoro}, P. and {Nu{\~n}ez Campos}, M. and {Delchambre}, L. and {Dell'Oro}, A. and {Esquej}, P. and {Fern{\'a}ndez-Hern{\'a}ndez}, J. and {Fraile}, E. and {Garabato}, D. and {Garc{\'\i}a-Lario}, P. and {Gosset}, E. and {Haigron}, R. and {Halbwachs}, J.-L. and {Hambly}, N.~C. and {Harrison}, D.~L. and {Hern{\'a}ndez}, J. and {Hestroffer}, D. and {Hodgkin}, S.~T. and {Holl}, B. and {Jan{\ss}en}, K. and {Jevardat de Fombelle}, G. and {Jordan}, S. and {Krone-Martins}, A. and {Lanzafame}, A.~C. and {L{\"o}ffler}, W. and {Marchal}, O. and {Marrese}, P.~M. and {Moitinho}, A. and {Muinonen}, K. and {Osborne}, P. and {Pancino}, E. and {Pauwels}, T. and {Recio-Blanco}, A. and {Reyl{\'e}}, C. and {Riello}, M. and {Rimoldini}, L. and {Roegiers}, T. and {Rybizki}, J. and {Sarro}, L.~M. and {Siopis}, C. and {Smith}, M. and {Sozzetti}, A. and {Utrilla}, E. and {van Leeuwen}, M. and {Abbas}, U. and {{\'A}brah{\'a}m}, P. and {Abreu Aramburu}, A. and {Aerts}, C. and {Aguado}, J.~J. and {Ajaj}, M. and {Aldea-Montero}, F. and {Altavilla}, G. and {{\'A}lvarez}, M.~A. and {Alves}, J. and {Anders}, F. and {Anderson}, R.~I. and {Anglada Varela}, E. and {Antoja}, T. and {Baines}, D. and {Baker}, S.~G. and {Balaguer-N{\'u}{\~n}ez}, L. and {Balbinot}, E. and {Balog}, Z. and {Barache}, C. and {Barbato}, D. and {Barros}, M. and {Barstow}, M.~A. and {Bartolom{\'e}}, S. and {Bassilana}, J.-L. and {Bauchet}, N. and {Becciani}, U. and {Bellazzini}, M. and {Berihuete}, A. and {Bernet}, M. and {Bertone}, S. and {Bianchi}, L. and {Binnenfeld}, A. and {Blanco-Cuaresma}, S. and {Blazere}, A. and {Boch}, T. and {Bombrun}, A. and {Bossini}, D. and {Bouquillon}, S. and {Bragaglia}, A. and {Bramante}, L. and {Breedt}, E. and {Bressan}, A. and {Brouillet}, N. and {Brugaletta}, E. and {Bucciarelli}, B. and {Burlacu}, A. and {Butkevich}, A.~G. and {Buzzi}, R. and {Caffau}, E. and {Cancelliere}, R. and {Cantat-Gaudin}, T. and {Carballo}, R. and {Carlucci}, T. and {Carnerero}, M.~I. and {Carrasco}, J.~M. and {Casamiquela}, L. and {Castellani}, M. and {Castro-Ginard}, A. and {Chaoul}, L. and {Charlot}, P. and {Chemin}, L. and {Chiaramida}, V. and {Chiavassa}, A. and {Chornay}, N. and {Comoretto}, G. and {Contursi}, G. and {Cooper}, W.~J. and {Cornez}, T. and {Cowell}, S. and {Crifo}, F. and {Cropper}, M. and {Crosta}, M. and {Crowley}, C. and {Dafonte}, C. and {Dapergolas}, A. and {David}, M. and {David}, P. and {de Laverny}, P. and {De Luise}, F. and {De March}, R.},
        title = "{Gaia Data Release 3. Summary of the content and survey properties}",
      journal = {\aap},
         year = 2023,
        month = jun,
       volume = {674},
          eid = {A1},
        pages = {A1},
          doi = {10.1051/0004-6361/202243940},
archivePrefix = {arXiv},
       eprint = {2208.00211},
 primaryClass = {astro-ph.GA},
       adsurl = {https://ui.adsabs.harvard.edu/abs/2023A&A...674A...1G}
}

@article{Foreman-Mackey2021,
       author = {{Foreman-Mackey}, Daniel and {Luger}, Rodrigo and {Agol}, Eric
                and {Barclay}, Thomas and {Bouma}, Luke G. and {Brandt},
                Timothy D. and {Czekala}, Ian and {David}, Trevor J. and
                {Dong}, Jiayin and {Gilbert}, Emily A. and {Gordon}, Tyler A.
                and {Hedges}, Christina and {Hey}, Daniel R. and {Morris},
                Brett M. and {Price-Whelan}, Adrian M. and {Savel}, Arjun B.},
        title = "{exoplanet: Gradient-based probabilistic inference for
                  exoplanet data \& other astronomical time series}",
      journal = {arXiv e-prints},
         year = 2021,
        month = may,
          eid = {arXiv:2105.01994},
        pages = {arXiv:2105.01994},
archivePrefix = {arXiv},
       eprint = {2105.01994},
 primaryClass = {astro-ph.IM},
       adsurl = {https://ui.adsabs.harvard.edu/abs/2021arXiv210501994F}
}

@ARTICLE{gelman1992,
       author = {{Gelman}, Andrew and {Rubin}, Donald B.},
        title = "{Inference from Iterative Simulation Using Multiple Sequences}",
      journal = {Statistical Science},
         year = 1992,
        month = jan,
       volume = {7},
        pages = {457-472},
          doi = {10.1214/ss/1177011136},
       adsurl = {https://ui.adsabs.harvard.edu/abs/1992StaSc...7..457G}
}

@ARTICLE{kwee1956,
       author = {{Kwee}, K.~K. and {van Woerden}, H.},
        title = "{A method for computing accurately the epoch of minimum of an eclipsing variable}",
      journal = {\bain},
         year = 1956,
        month = may,
       volume = {12},
        pages = {327},
       adsurl = {https://ui.adsabs.harvard.edu/abs/1956BAN....12..327K}
}

@ARTICLE{lindegren2021,
       author = {{Lindegren}, L. and {Bastian}, U. and {Biermann}, M. and {Bombrun}, A. and {de Torres}, A. and {Gerlach}, E. and {Geyer}, R. and {Hern{\'a}ndez}, J. and {Hilger}, T. and {Hobbs}, D. and {Klioner}, S.~A. and {Lammers}, U. and {McMillan}, P.~J. and {Ramos-Lerate}, M. and {Steidelm{\"u}ller}, H. and {Stephenson}, C.~A. and {van Leeuwen}, F.},
        title = "{Gaia Early Data Release 3. Parallax bias versus magnitude, colour, and position}",
      journal = {\aap},
         year = 2021,
        month = may,
       volume = {649},
          eid = {A4},
        pages = {A4},
          doi = {10.1051/0004-6361/202039653},
archivePrefix = {arXiv},
       eprint = {2012.01742},
 primaryClass = {astro-ph.IM},
       adsurl = {https://ui.adsabs.harvard.edu/abs/2021A&A...649A...4L}
}

@ARTICLE{hoffman2011,
       author = {{Hoffman}, Matthew D. and {Gelman}, Andrew},
        title = "{The No-U-Turn Sampler: Adaptively Setting Path Lengths in Hamiltonian Monte Carlo}",
      journal = {arXiv e-prints},
         year = 2011,
        month = nov,
          eid = {arXiv:1111.4246},
        pages = {arXiv:1111.4246},
          doi = {10.48550/arXiv.1111.4246},
archivePrefix = {arXiv},
       eprint = {1111.4246},
 primaryClass = {stat.CO},
       adsurl = {https://ui.adsabs.harvard.edu/abs/2011arXiv1111.4246H}
}

@ARTICLE{irwin1952,
       author = {{Irwin}, John B.},
        title = "{The Determination of a Light-Time Orbit.}",
      journal = {\apj},
         year = 1952,
        month = jul,
       volume = {116},
        pages = {211},
          doi = {10.1086/145604},
       adsurl = {https://ui.adsabs.harvard.edu/abs/1952ApJ...116..211I}
}

@INPROCEEDINGS{xtrema2015,
       author = {{Bahar}, E. and {{\c{S}}enavc{\i}}, H.~V. and {Ba{\c{s}}t{\"u}rk}, {\"O}.},
        title = "{A Python-based GUI Software to Calculate Times of Maximum and Minimum: Xtrema}",
    booktitle = {Living Together: Planets, Host Stars and Binaries},
         year = 2015,
       editor = {{Rucinski}, S.~M. and {Torres}, G. and {Zejda}, M.},
       series = {Astronomical Society of the Pacific Conference Series},
       volume = {496},
        month = jul,
        pages = {288},
       adsurl = {https://ui.adsabs.harvard.edu/abs/2015ASPC..496..288B}
}

@ARTICLE{muterspaugh2006,
       author = {{Muterspaugh}, M.~W. and {Lane}, B.~F. and {Konacki}, M. and {Burke}, B.~F. and {Colavita}, M.~M. and {Kulkarni}, S.~R. and {Shao}, M.},
        title = "{PHASES differential astrometry and the mutual inclination of the V819 Herculis triple star system}",
      journal = {\aap},
         year = 2006,
        month = feb,
       volume = {446},
       number = {2},
        pages = {723-732},
          doi = {10.1051/0004-6361:20053749},
archivePrefix = {arXiv},
       eprint = {astro-ph/0509855},
 primaryClass = {astro-ph},
       adsurl = {https://ui.adsabs.harvard.edu/abs/2006A&A...446..723M}
}

@ARTICLE{muterspaugh2008,
       author = {{Muterspaugh}, Matthew W. and {Lane}, Benjamin F. and {Fekel}, Francis C. and {Konacki}, Maciej and {Burke}, Bernard F. and {Kulkarni}, S.~R. and {Colavita}, M.~M. and {Shao}, M. and {Wiktorowicz}, Sloane J.},
        title = "{Masses, Luminosities, and Orbital Coplanarities of the {\ensuremath{\mu}} Orionis Quadruple-Star System from Phases Differential Astrometry}",
      journal = {\aj},
         year = 2008,
        month = mar,
       volume = {135},
       number = {3},
        pages = {766-776},
          doi = {10.1088/0004-6256/135/3/766},
archivePrefix = {arXiv},
       eprint = {0710.2126},
 primaryClass = {astro-ph},
       adsurl = {https://ui.adsabs.harvard.edu/abs/2008AJ....135..766M}
}

@ARTICLE{obrien2011,
       author = {{O'Brien}, David P. and {McAlister}, Harold A. and {Raghavan}, Deepak and {Boyajian}, Tabetha S. and {ten Brummelaar}, Theo A. and {Sturmann}, Judit and {Sturmann}, Laszlo and {Turner}, Nils H. and {Ridgway}, Stephen},
        title = "{Inner Orbits in Hierarchical Triple Systems from the CHARA Array. I. V819 Her B}",
      journal = {\apj},
         year = 2011,
        month = feb,
       volume = {728},
       number = {2},
          eid = {111},
        pages = {111},
          doi = {10.1088/0004-637X/728/2/111},
       adsurl = {https://ui.adsabs.harvard.edu/abs/2011ApJ...728..111O}
}

@article{pymc2023,
  title = {{PyMC}: A Modern and Comprehensive Probabilistic Programming Framework in {P}ython},
  author = {Oriol Abril-Pla and Virgile Andreani and Colin Carroll and Larry Dong and Christopher J. Fonnesbeck and Maxim Kochurov and Ravin Kumar and Junpeng Lao and Christian C. Luhmann and Osvaldo A. Martin and Michael Osthege and Ricardo Vieira and Thomas Wiecki and Robert Zinkov },
  journal = {{PeerJ} Computer Science},
  volume = {9},
  number = {e1516},
  doi = {10.7717/peerj-cs.1516},
  year = {2023}
}

@ARTICLE{scarfe1994,
       author = {{Scarfe}, C.~D. and {Barlow}, D.~J. and {Fekel}, F.~C. and {Rees}, R.~F. and {Lyons}, R.~W. and {Bolton}, C.~T. and {McAlister}, H.~A. and {Hartkopf}, W.~I.},
        title = "{The Spectroscopic-Speckle Triple System HR 6469}",
      journal = {\aj},
         year = 1994,
        month = apr,
       volume = {107},
        pages = {1529},
          doi = {10.1086/116963},
       adsurl = {https://ui.adsabs.harvard.edu/abs/1994AJ....107.1529S}
}

@ARTICLE{vanHamme1994,
       author = {{van Hamme}, W.~V. and {Hall}, D.~S. and {Hargrove}, A.~W. and {Henry}, G.~W. and {Wasson}, R. and {Barkslade}, W.~S. and {Chang}, S. and {Fried}, R.~E. and {Green}, C.~L. and {Lines}, H.~C. and {Lines}, R.~D. and {Nielsen}, P. and {Powell}, H.~D. and {Reisenweber}, R.~C. and {Rogers}, C.~W. and {Shervais}, S. and {Tatum}, R.},
        title = "{The Two Variables in The Triple System HR 6469=V819 Her: One Eclipsing, One Spotted}",
      journal = {\aj},
         year = 1994,
        month = apr,
       volume = {107},
        pages = {1521},
          doi = {10.1086/116962},
       adsurl = {https://ui.adsabs.harvard.edu/abs/1994AJ....107.1521V}
}

@ARTICLE{wasson1994,
       author = {{Wasson}, Rick and {Hall}, Douglas S. and {Hargrove}, Adam W. and {Bertoglio}, Alessandro and {Burke}, Jr., Edward W. and {Frey}, Gary J. and {Fried}, Robert E. and {Galli}, Gianni and {Green}, Charles L. and {Iacovone}, Robert A., Nicola Jones and {Lines}, Helen C. and {Lines}, Richard D. and {Mattingly}, Philip and {Milton}, Russell E. and {Nelson}, Robert and {Nolthenius}, Richard and {Powell}, Harry D. and {Nix}, Russell M. and {Passour}, John L. and {Robb}, Steve and {Snyder}, Leroy F. and {Williams}, David B.},
        title = "{Ten Years of Eclipse Timings to Refine The Period of HR 6469=V819 HER}",
      journal = {\aj},
         year = 1994,
        month = apr,
       volume = {107},
        pages = {1514},
          doi = {10.1086/116961},
       adsurl = {https://ui.adsabs.harvard.edu/abs/1994AJ....107.1514W}
}

@ARTICLE{wilsondev1971,
       author = {{Wilson}, Robert E. and {Devinney}, Edward J.},
        title = "{Realization of Accurate Close-Binary Light Curves: Application to MR Cygni}",
      journal = {\apj},
         year = 1971,
        month = jun,
       volume = {166},
        pages = {605},
          doi = {10.1086/150986},
       adsurl = {https://ui.adsabs.harvard.edu/abs/1971ApJ...166..605W}
}

@ARTICLE{wilsondev1973,
       author = {{Wilson}, R.~E. and {Devinney}, E.~J.},
        title = "{Fundamental data for contact binaries: RZ Comae Berenices, RZ Tauri, and AW Ursae Majoris.}",
      journal = {\apj},
         year = 1973,
        month = jun,
       volume = {182},
        pages = {539},
          doi = {10.1086/152162},
       adsurl = {https://ui.adsabs.harvard.edu/abs/1973ApJ...182..539W}
}

@ARTICLE{zasche2014,
       author = {{Zasche}, P. and {Uhl{\`a}{\v{r}}}, R. and {Svoboda}, P.},
        title = "{The Study of Triple Systems V819 Her, V2388 Oph, and V1031 Ori}",
      journal = {\actaa},
         year = 2014,
        month = jun,
       volume = {64},
       number = {2},
        pages = {125-150},
          doi = {10.48550/arXiv.1407.5741},
archivePrefix = {arXiv},
       eprint = {1407.5741},
 primaryClass = {astro-ph.SR},
       adsurl = {https://ui.adsabs.harvard.edu/abs/2014AcA....64..125Z}
}

@ARTICLE{lindegren2018,
  author  = {Lindegren, L. and Hern\'andez, J. and Bombrun, A. and
             Klioner, S. and Bastian, U. and Ramos-Lerate, M. and
             {de Torres}, A. and Steigelm\"uller, H. and Stephenson, C.
             and Hobbs, D. and others},
  title   = {{Gaia Data Release 2. The astrometric solution}},
  journal = {\aap},
  year    = {2018},
  volume  = {616},
  pages   = {A2},
  doi     = {10.1051/0004-6361/201832727}
}

@ARTICLE{stassun2018,
  author  = {Stassun, Keivan G. and Torres, Guillermo},
  title   = {{Evidence for a Systematic Offset of $-80\,\mu$as in
             the Gaia DR2 Parallaxes}},
  journal = {\apj},
  year    = {2018},
  volume  = {862},
  pages   = {61},
  doi     = {10.3847/1538-4357/aacafc}
}

@ARTICLE{zinn2019,
  author  = {Zinn, Joel C. and Pinsonneault, Marc H. and Huber, Daniel
             and Stello, Dennis},
  title   = {{Confirmation of the Gaia DR2 Parallax Zero-point Offset
             Using Asteroseismology and Spectroscopy in the Kepler
             Field}},
  journal = {\apj},
  year    = {2019},
  volume  = {878},
  pages   = {136},
  doi     = {10.3847/1538-4357/ab1f66}
}

@ARTICLE{vanleeuwen2007,
  author  = {{van Leeuwen}, Floor},
  title   = {{Validation of the new Hipparcos reduction}},
  journal = {\aap},
  year    = {2007},
  volume  = {474},
  pages   = {653--664},
  doi     = {10.1051/0004-6361:20078357}
}

@ARTICLE{groenewegen2023,
       author = {{Groenewegen}, M.~A.~T.},
        title = "{Orbital parallax of binary systems compared to Gaia DR3 and the parallax zero-point offset at bright magnitudes}",
      journal = {\aap},
         year = 2023,
        month = jan,
       volume = {669},
          eid = {A4},
        pages = {A4},
          doi = {10.1051/0004-6361/202244479},
archivePrefix = {arXiv},
       eprint = {2210.14734}
}

@ARTICLE{howell2024,
       author = {{Howell}, Steve B. and {Martinez}, Arturo O. and {Hope}, Douglas A. and {Ciardi}, David R. and {Jefferies}, Stuart M. and {Baron}, Fabien R. and {Lund}, Michael B.},
        title = "{High-contrast, High-angular-resolution Optical Speckle Imaging: Uncovering Hidden Stellar Companions}",
      journal = {\aj},
         year = 2024,
        month = jun,
       volume = {167},
       number = {6},
          eid = {258},
        pages = {258},
          doi = {10.3847/1538-3881/ad3df2},
archivePrefix = {arXiv},
       eprint = {2404.15528}
}

@ARTICLE{kipping2013,
       author = {{Kipping}, David M.},
        title = "{Parametrizing the exoplanet eccentricity distribution with the beta distribution}",
      journal = {\mnras},
         year = 2013,
       volume = {434},
        pages = {L51-L55},
          doi = {10.1093/mnrasl/slt075}
}

@ARTICLE{muterspaugh2010,
       author = {{Muterspaugh}, Matthew W. and {Lane}, Benjamin F. and {Kulkarni}, S.~R. and {Konacki}, Maciej and {Burke}, Bernard F. and {Colavita}, M.~M. and {Shao}, M. and {Wiktorowicz}, Sloane J. and {O'Connell}, J.},
        title = "{The PHASES Differential Astrometry Data Archive. I. Measurements and Description}",
      journal = {\aj},
         year = 2010,
       volume = {140},
        pages = {1579-1622},
archivePrefix = {arXiv},
       eprint = {1010.4041}
}

@ARTICLE{horch2021,
       author = {{Horch}, Elliott P. and {Broderick}, Kyle G. and {Casetti-Dinescu}, Dana I. and {Henry}, Todd J. and {Fekel}, Francis C. and {Muterspaugh}, Matthew W. and {Willmarth}, Daryl W. and {Winters}, Jennifer G. and {van Belle}, Gerard T. and {Clark}, Catherine A. and {Everett}, Mark E.},
        title = "{Observations with the Differential Speckle Survey Instrument. X. Preliminary Orbits of K-dwarf Binaries and Other Stars}",
      journal = {\aj},
         year = 2021,
        month = jun,
       volume = {161},
          eid = {295},
        pages = {295},
archivePrefix = {arXiv},
       eprint = {2104.07760}
}

\appendix

\section{Posterior Distributions and Parameter Correlations}
\label{sec:appendix_plots}

\begin{figure*}
    \centering
    \includegraphics[width=0.90\textwidth]{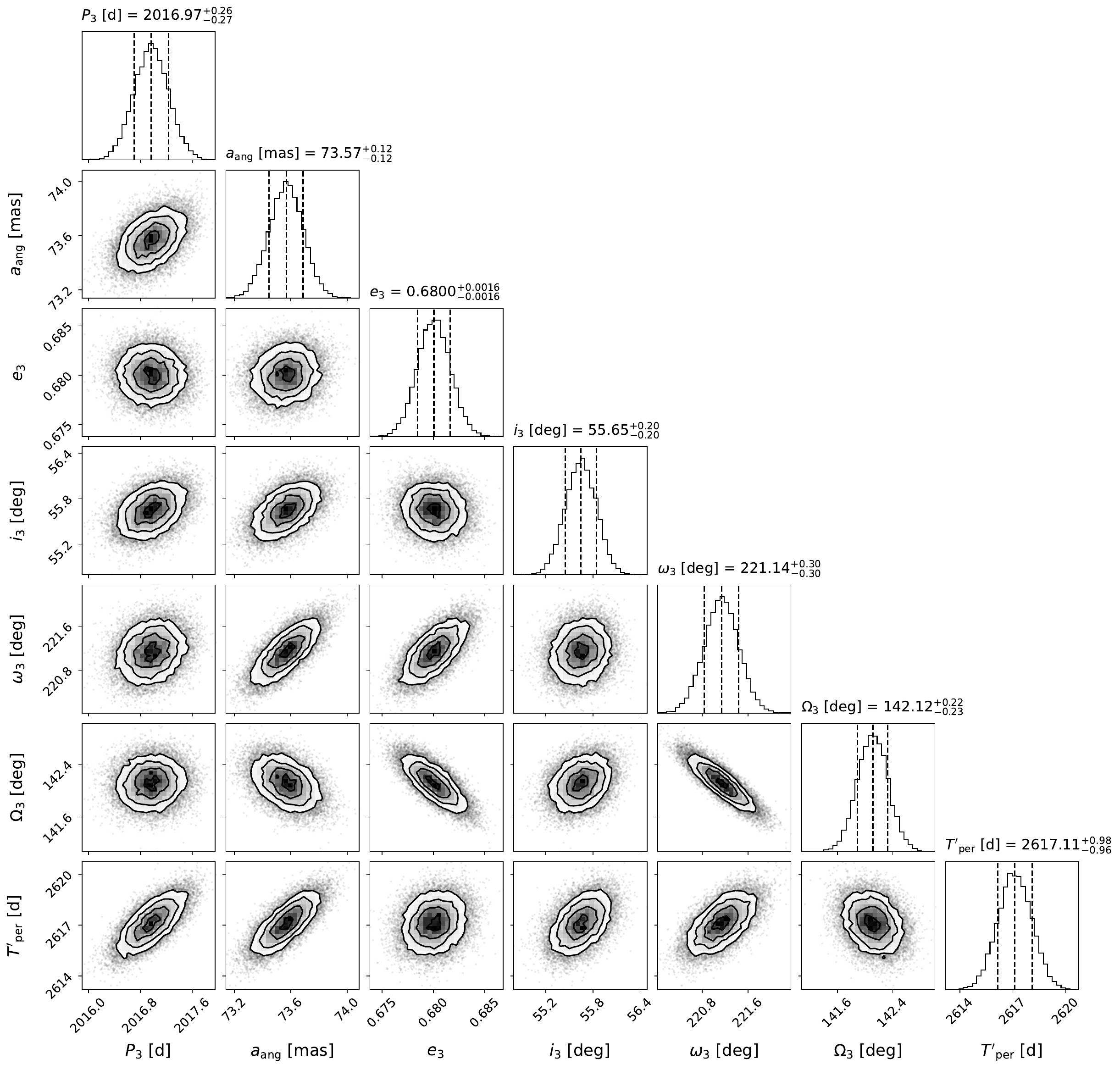}
    \caption{Posterior probability distributions and two-dimensional correlation contours for the primary sampled orbital elements of the V819 Herculis outer orbit. The dashed vertical lines indicate the posterior median and the \new{16th and 84th} percentiles.}
    \label{fig:corner_orbital}
\end{figure*}

\begin{figure*}
    \centering
    \includegraphics[width=0.90\textwidth]{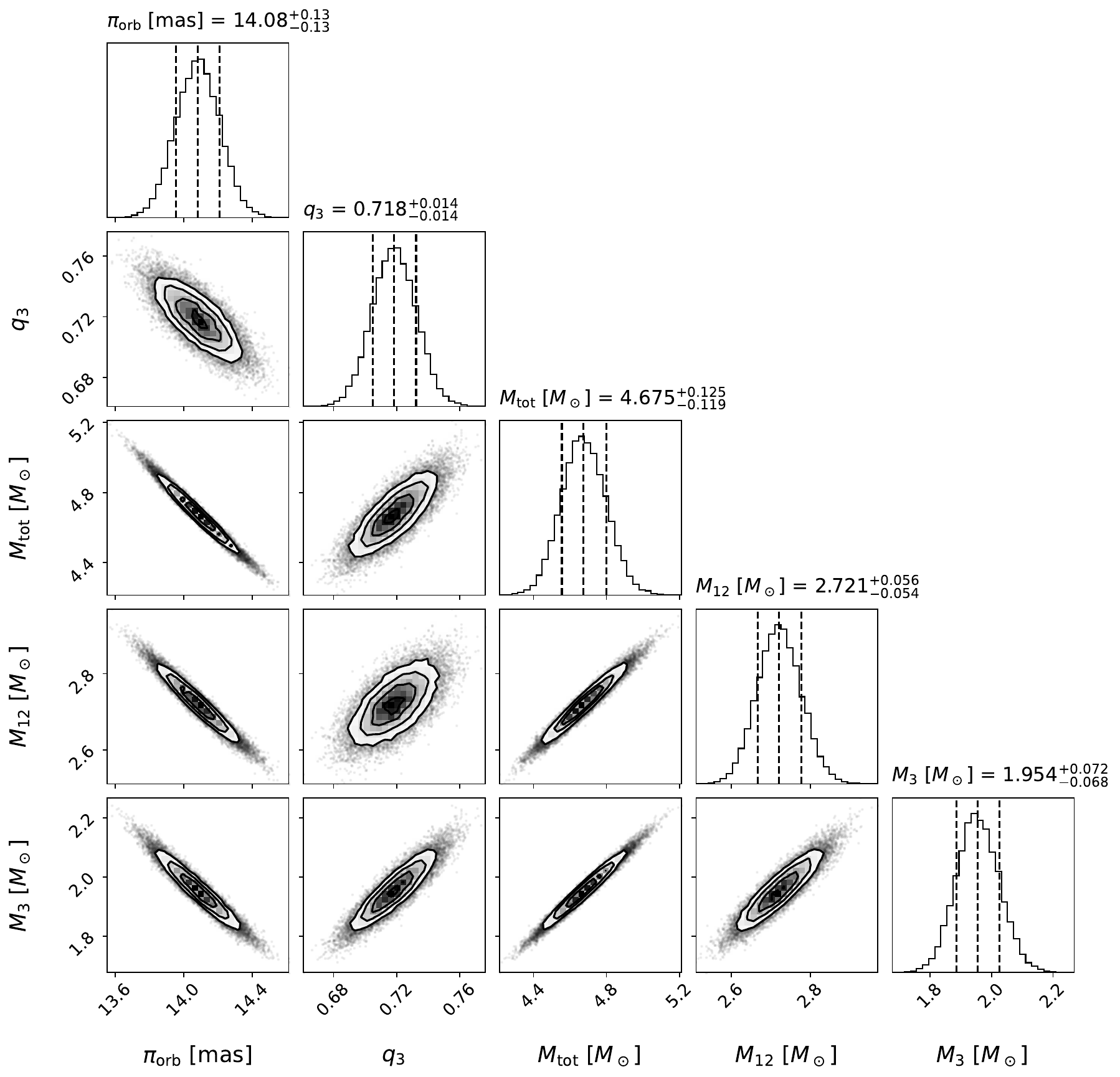}
    \caption{\new{Posterior probability distributions and two-dimensional correlations of the orbital parallax, the mass ratio $q_3 = M_3/M_{12}$ and the component masses. The strong anticorrelation between $\pi_{\rm orb}$ and $M_{\rm tot}$ reflects $M_{\rm tot} \propto \pi_{\rm orb}^{-3}$. The dashed vertical lines indicate the posterior median and the 16th and 84th percentiles.}}
    \label{fig:corner_physical}
\end{figure*}

In this section, we present the complete corner plots demonstrating the convergence and parameter degeneracies of our joint MCMC analysis. Figure~\ref{fig:corner_orbital} illustrates the posterior distributions for the primary sampled orbital parameters, including the periods, angular scales, and orientations. Figure~\ref{fig:corner_physical} displays \new{the orbital parallax, the mass ratio and the component masses, illustrating how the dynamical masses follow from the joint solution}.

\bsp
\label{lastpage}
\end{document}